\documentclass[journal abbreviation]{copernicus}
\begin{document}

%\title{Galactic Cosmic Rays and their Interaction with the Heliosphere}
\title{Cosmic Rays and Stochastic Magnetic Reconnection in the Heliotail}

\author[1,2]{P. Desiati}
\author[2]{A. Lazarian}

\affil[1]{Wisconsin IceCube Particle Astrophysics Center, University of Wisconsin, Madison, WI 53706, U.S.A.}
\affil[2]{Dept. of Astronomy, University of Wisconsin, Madison, WI 53706, U.S.A.}

%% The [] brackets identify the author to the corresponding affiliation, 1, 2, 3, etc. should be inserted.

\runningtitle{Cosmic Rays and Reconnection in the Heliotail}

\runningauthor{Desiati \& Lazarian}

\correspondence{P. Desiati (desiati@icecube.wisc.edu)\\A. Lazarian (lazarian@astro.wisc.edu)}

\received{}
\pubdiscuss{} %% only important for two-stage journals
\revised{}
\accepted{}
\published{}

%% These dates will be inserted by the Publication Production Office during the typesetting process.

\firstpage{1}
\maketitle

\begin{abstract}
Galactic cosmic rays are believed to be generated by diffusive shock acceleration processes in Supernova Remnants, and the arrival direction is likely determined by the distribution of their sources throughout the Galaxy, in particular by the nearest and youngest ones. Transport to Earth through the interstellar medium is expected to affect the cosmic ray properties as well. However, the observed anisotropy of TeV cosmic rays and its energy dependence cannot be explained with diffusion models of particle propagation in the Galaxy. Within a distance of a few parsec, diffusion regime is not valid and particles with energy below about 100 TeV must be influenced by the heliosphere and its elongated tail. The observation of a highly significant localized excess region of cosmic rays from the apparent direction of the downstream interstellar flow at 1-10 TeV energies might provide the first experimental evidence that the heliotail can affect the transport of energetic particles. In particular, TeV cosmic rays propagating through the heliotail interact with the 100-300 AU wide magnetic field polarity domains generated by the 11 year cycles. Since the strength of non-linear convective processes is expected to be larger than viscous damping, the plasma in the heliotail is turbulent. Where magnetic field domains converge on each other due to solar wind gradient, stochastic magnetic reconnection likely occurs. Such processes may be efficient enough to re-accelerate a fraction of TeV particles as long as scattering processes are not strong. Therefore the fractional excess of TeV cosmic rays from the narrow region toward the heliotail direction traces sightlines with the lowest smearing scattering effects, that can also explain the observation of a harder than average energy spectrum.
\end{abstract}

\introduction
%% \introduction[modified heading if necessary]
\label{sec:intro}

\begin{figure}[!t]
\center{
\includegraphics[width=\columnwidth]{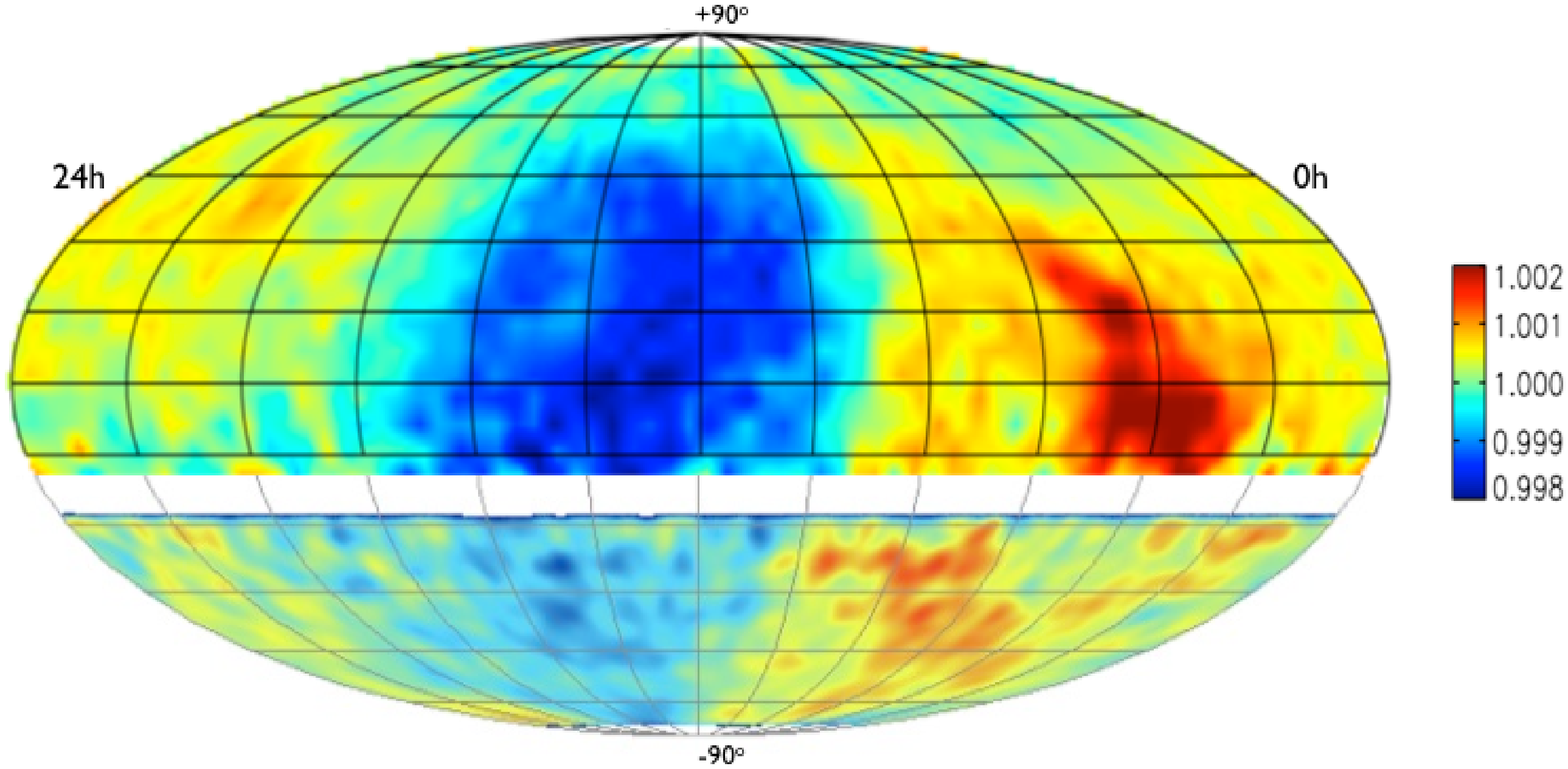}
\includegraphics[width=0.9\columnwidth]{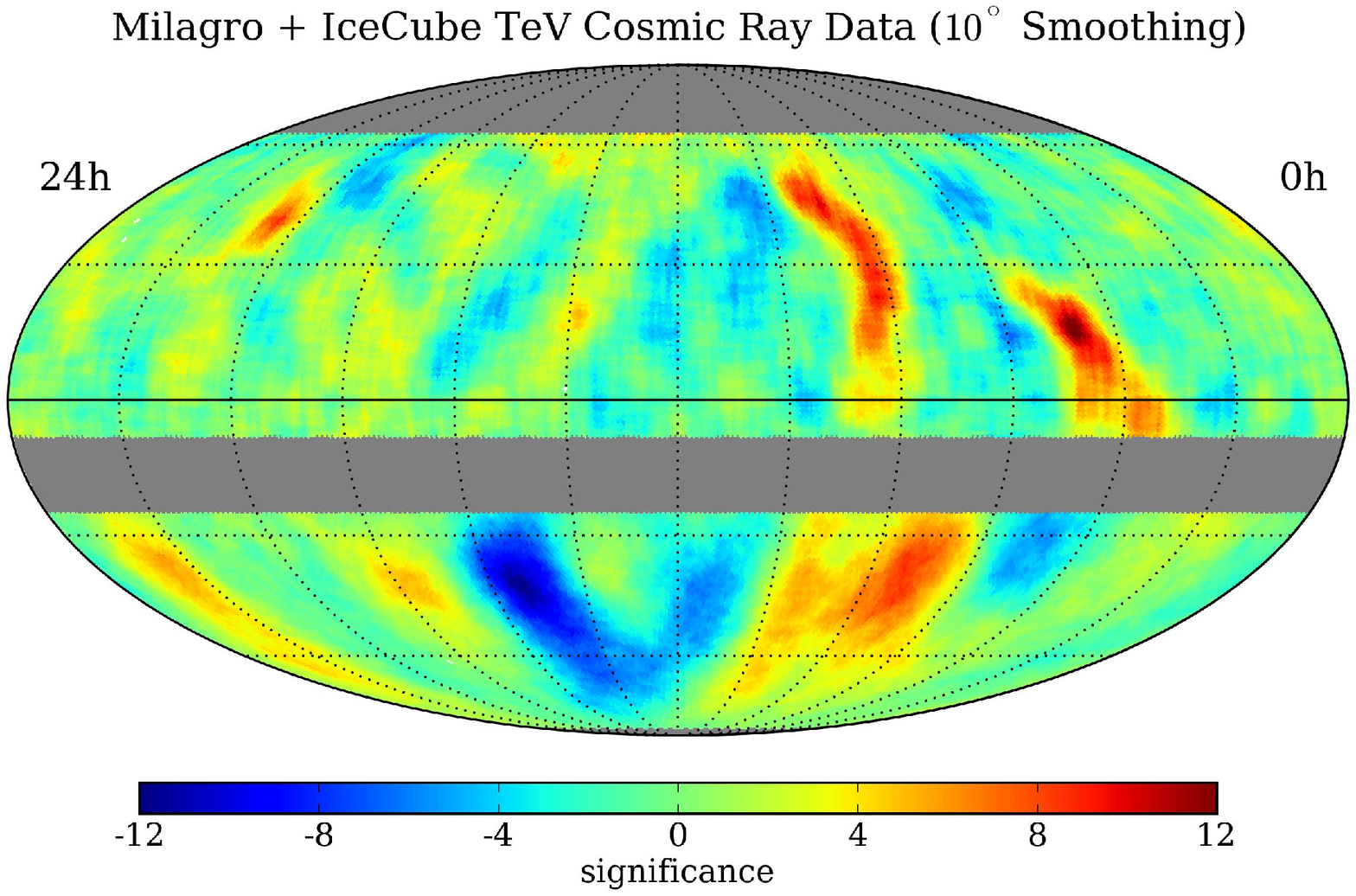}
}
\caption{{\it Top panel}: Map in equatorial coordinates of the relative intensity of the cosmic ray arrival distribution as observed by the Tibet AS$\gamma$ at about 5 TeV in the northern hemisphere~(from \cite{amenomori11}) and by the IceCube Observatory at a median energy of 20 TeV in the southern hemisphere~(from~\cite{abbasi}).  {\it Bottom panel}: Map in equatorial coordinates of the statistical significance of the cosmic ray arrival direction distribution as observed by Milagro at about 1 TeV in the northern hemisphere~(from~\cite{abdo2}) and by the IceCube Observatory at a median energy of 20 TeV in the southern hemisphere (from~\cite{abbasi11b}). In this map features with angular extension larger than $30^{\circ}-60^{\circ}$ are filtered out.}
\label{fig:anisotropy}
\end{figure}

During the last decades, galactic cosmic rays have been found to have a small but measurable energy dependent uneven arrival direction distribution, with a relative amplitude of order $10^{-4}-10^{-3}$. This anisotropy was observed in the northern hemisphere from energies of tens to several hundreds GeV with muon detectors~\citep{nagashima,munakata11}, and in the multi-TeV energy range with Tibet AS$\gamma$ array~\citep{amenomori,amenomori11}, Super-Kamiokande \citep{guillian}, Milagro \citep{abdo} and ARGO-YBJ~\citep{argo,shuwang11}. An anisotropy was also observed at an energy in excess of about 100 TeV with the EAS-TOP shower array~\citep{aglietta}. Recently similar observations were reported in the southern hemisphere at energies of 10's to 100's TeV with the IceCube Observatory~\citep{abbasi,abbasi11}. While at $\sim$ 10 TeV the anisotropy appears to be topologically connected to the GeV-TeV observations in the north, above about 100 TeV the global anisotropy persists but with a different phase, consistently with the results from~\cite{aglietta}. The top panel of Fig.~\ref{fig:anisotropy} shows the combined map in equatorial coordinates of relative intensity of cosmic ray arrival direction distribution observed by Tibet AS$\gamma$ at about 5 TeV in the northern hemisphere, and by IceCube at about 20 TeV in the southern hemisphere.

The arrival distribution of sub-TeV cosmic rays revealed the existence of two kinds of anisotropies, as discussed in~\cite{nagashima}. One is a global non-dipolar anisotropy with relative excess approximately centered around equatorial right ascension of 0 hr, and an increasing amplitude up to 1-10 TeV energies. The other is a directional excess region confined in a cone of half opening angle of 68$^{\circ}$ from right ascension of about 6 hr, and observed for energies below TeV. This region covers a portion of the sky that includes the direction of the heliospheric tail (or heliotail), which is the region of the heliosphere downstream the interstellar wind delimited within the heliopause, i.e. the boundary that separates the solar wind and interstellar plasmas~\citep{izmodenov}. Its origin was therefore attributed to some unidentified anisotropic process occurring in the heliotail, and thus it was called "tail-in" excess.

At the higher TeV energies, while the tail-in broad excess becomes sub-dominant, the global anisotropy shows evidence of statistically significant small angular structures from the same direction in the sky. In particular, using experimental techniques in the attempt to isolate relatively localized excess or deficit regions of events that overlap over the smooth global anisotropy, angular features of order 20$^{\circ}$-30$^{\circ}$ were discovered. Two separate highly significant localized fractional excess regions of cosmic rays were reported in the northern hemisphere by Milagro~\citep{abdo2}, and also by Tibet AS$\gamma$~\citep{amenomori2} and ARGO-YBJ~\citep{vernetto,iuppa11}. The observation of a small scale anisotropy at multi-TeV energies was reported by IceCube, in the southern hemisphere, as well~\citep{abbasi11b}. The bottom panel of Fig.~\ref{fig:anisotropy} shows the combined map in equatorial coordinates of statistical significance of the cosmic ray arrival direction distribution where only features with angular extension smaller than about 60$^{\circ}$ are visible. Such small scale features lay in the same portion of the sky where the tail-in excess was dominant at lower energy, especially the one toward the heliotail direction with equatorial coordinates ($\alpha$, $\delta$) $\approx$ (5 hr, +17$^{\circ}$). %Such a discovery stimulated astrophysical interpretations that would link the anisotropy to nearby sources of cosmic rays or to effects of propagation through turbulent interstellar magnetic field. %such as the progenitor supernova that gave birth to Geminga pulsar 350,000 years ago at a distance of order 100 pc~\citep{salvati,drury,salvati2}. Unfortunately nothing or very little is known about the local interstellar medium properties, therefore cosmic ray diffusion is not sufficiently constraint to provide a coherent scenario that can explain the observations without considerable fine tuning.

%\begin{figure}[!t]
%\includegraphics[width=\columnwidth]{figs/Low_energy_muons.eps}
%\caption{Contours of the values of relative intensity in the arrival direction distribution of cosmic rays at 500 GeV. The solid contours indicate the region of tail-in excess and the dashed contours indicate the loss cone deficit region. The bold dotted line is the galactic equator. The contours range between 0.072\% and -0.077\% in steps of 0.006\% (from~\cite{hall99}).}
%\label{fig:lowenergy}
%\end{figure}

At an energy in excess of about 100 TeV, where the anisotropy has a different topology than at lower energy, cosmic ray particles are hardly influenced by the heliosphere and its elongated tail, and their arrival direction might hold information on the Local Interstellar Magnetic Field (LIMF) on a larger scale. If the extended heliotail induces a significant perturbation in the local interstellar medium, that can affect the arrival direction of multi-TeV cosmic ray particles, then the anisotropy can be considered as an indirect probe of how the LIMF influences the heliospheric boundary itself (see~\cite{locscat}). Moreover, cosmic rays below about 10 TeV are expected to be influenced by magnetic fields inside the heliotail as well. The concurrent effects of magnetic reconnection and scattering processes might be able to explain some observations, although more experimental results and further developments in heliospheric Magneto Hydro-Dynamic (MHD) simulations are needed for better constrain models.

The origin of the cosmic ray anisotropy, its persistence in a wide energy range and its angular structure, is currently subject of debate. In this paper we will briefly report the interpretations provided by various authors (in Sec.~\ref{sec:interp}), with an emphasis on a possible phenomenological connection between the broad tail-in excess of sub-TeV cosmic rays and the localized fractional excess of multi-TeV cosmic rays from the direction of the heliotail. We'll then describe the magnetic field structure in the heliotail as shaped by solar cycles and rotation in Sec.~\ref{sec:magfi}. An overview on magnetic reconnection processes is given in Sec.~\ref{sec:rec}, with an emphasis on stochastic reconnection, assumed to contribute to the origin of the anomalies observed toward the heliotail. Sec.~\ref{sec:acc} addresses the proposed mechanism of cosmic ray re-acceleration in the heliotail and its effective influence in relation to scattering processes.

%It is proposed that both sub-TeV tail-in excess and the multi-TeV localized excess of cosmic rays might be caused by magnetic reconnection in the heliosphere and, in particular in the heliotail, where the distance scale might be long enough to induce sufficient acceleration at high energy. The very idea of appealing to magnetic reconnection for the acceleration of energetic particles can be traced back to pioneeding works by Giovanelli (1946) and Dungey (1953). The uncertainties with understanding of fast reconnection were one of the impediments for applying the process to energetic particle acceleration (see Lazarian \& Opher 2009). We appeal to the model of reconnection of weakly stochastic field in Lazarian \& Vishniac (1999), which was identified as a cause of First Order Fermi acceleration (see de Gouveia dal Pino \& Lazarian 2005, Lazarian 2005).

%In what follows we present the observational evidence for the existence of the cosmic ray excess in the direction of the solar system magnetotail in \S \ref{sec:obs}, discuss existing explanations of this excess in \S \ref{sec:interp}. The structure of the magnatotail with magnetic field reversals arising from the solar cycle is presented in \S \ref{sec:magfi} and the mechanism of acceleration of cosmic rays in the magnetotail is outlined in \S \ref{sec:magrec}. The discussion of the results and a short summary are given by \S \ref{sec:disc} and \S \ref{sec:summ}, respectively. 

%**********************************************************************************************************************************
\section{Cosmic Rays Anisotropy}
\label{sec:interp}

%----------------------------------------------------------------------------- to section 2
%The second most significant localized fractional excess region discovered in the northern hemisphere (the so-called region B in~\cite{abdo2}), has an elongated structure possibly connected to an arc-like feature observed in the southern hemisphere~\citep{abbasi11b}. This arc of TeV cosmic rays is located relatively close to the sightlines perpendicular to the local interstellar magnetic field as inferred from a number of observations~\citep{locscat}.
%
%The most significant of the two excess regions (the so-called region A in~\cite{abdo2}) coincides with the direction of the heliotail (or better the downstream direction of the interstellar medium wind). Given its small angular scale, it is argued in this paper that its origin is likely related to local processes as well, i.e. within the cosmic particles free mean path in the local interstellar medium. In particular that the broad tail-in excess of sub-TeV cosmic rays and the localized fractional excess of multi-TeV cosmic rays from the direction of the heliotail, are two manifestations of the same phenomenology where the different appearance is due solely to the different energy scale. A fraction of cosmic rays propagating through the heliotail may undergo re-acceleration via stochastic magnetic reconnection processes that occur between the sectored heliospheric magnetic field reversal regions that result from the 11 year solar cycles~\citep{reconnection}.
%-----------------------------------------------------------------------------

The origin of cosmic ray anisotropy is still unknown. The relative motion of the solar system with respect to the cosmic ray plasma rest frame (for instance due to galactic rotation) would produce a dipolar anisotropy in the direction of the motion~\citep{compton,compton2}. Such Compton-Getting effect was not singled out from observations yet, inducing to a possible conclusion that the bulk of galactic cosmic rays co-rotates with the solar system~\citep{amenomori}. Moreover, in the scenario where galactic cosmic rays are accelerated in supernova remnants, their arrival direction should have a relative excess toward the galactic center, i.e. the line of sight with the larger expected number of sources. On the other hand the nearest galactic sources would dominate the observed arrival distribution, and changes in anisotropy amplitude and phase with cosmic ray energy can arise as a natural consequence of the stochastic nature of their sources in the local interstellar medium~\citep{erlykin, blasi}.

Propagation properties of cosmic rays in the interstellar medium are likely to have an important role in shaping the anisotropy as well~\citep{battaner}. For instance, a scenario where the large scale anisotropy is linked to diffusion of cosmic rays through the LIMF connecting the solar system to the interstellar medium outside the local interstellar cloud (where the solar system currently resides), was proposed by~\cite{amenomori2,amenomori11b}. This model accounts for the apparent quadrupolar contribution observed with the large scale anisotropy. In~\cite{frisch2011b} it is noted that the tail-in excess region, besides including the heliotail direction, is centered around the direction of the LIMF, therefore linking its origin to their propagation deep inside the tail or to streaming along the LIMF or the S1 sub-shell of Loop I superbubble.

Within a distance of a few times the mean free path, diffusion regime breaks down and propagation of cosmic rays depends on their interaction with the turbulence ripples of the LIMF. Even though observations suggest that the LIMF is coherent over scales of about 100 pc, they also imply variations in field directions of less than 30$^{\circ}$ - 40$^{\circ}$, that can be attributed to turbulence~\citep{frisch2011b}. Scattering of TeV-PeV cosmic ray particles with the turbulent interstellar magnetic field within the mean free path (i.e. a few 10's pc) can generate intermediate and small scale perturbations over an underlying large scale anisotropy~\citep{giacinti}. The observed anisotropy structure, therefore, could be used to infer turbulence properties of the LIMF. At energies below about 100 TeV the proton gyro-radius is a few thousands AU, thought to be comparable to the length of the heliotail~\citep{izmodenov03}. At these energies cosmic ray anisotropy is likely influenced by the extended and turbulent heliospheric magnetic field, and localized features in arrival direction can arise from the scattering of energetic cosmic ray particles with the heliospheric magnetic field ordered by the LIMF direction~\citep{locscat}.

%However, the combined study of the energy evolution of the anisotropy and of its angular scale structure seem to suggest that the observation might likely be generated by a combination of effects, caused by phenomenologies at different distance scales from Earth, and impacting at different energy scales. At the same time, some features observed at different energies and apparently uncorrelated, could also have the same origin.
Another model aimed to explain the origin of the TeV small scale anisotropic features, appeals to the observation that the two localized excess regions in the northern hemisphere are seemingly close to the so-called Hydrogen Deflection Plane (HDP), which is the plane containing the directions of the interstellar flow and of the magnetic field upstream the heliospheric nose ~\citep{amenomori11b}. According to this model cosmic rays propagating along the heliotail within the HDP are bent by the heliospheric magnetic field so that two localized excess regions are formed symmetrically separated with respect to the direction of the heliotail on the HDP. This implies that the heliospheric magnetic field between about 70 AU and 340 AU along the heliotail is responsible for the two localized regions observed in the northern hemisphere in the energy range between 4 and 30 TeV. The heliospheric magnetic field has a complex structure determined by the combined effects of the 26 day rotation period of the Sun and of the 11 year solar cycle~\citep{pogorelov}. This complex time-dependent magnetic field structure should produce an observable time variability in the relative intensity and position of the localized fractional regions over an 11 year period.

Some other models rely on an astrophysical origin of the observation. In \cite{salvati, drury, salvati2} it is noted that the two observed localized excess regions in the northern hemisphere, surround the present day apparent location of Geminga pulsar. The supernova that gave birth to the pulsar exploded about 340,000 years ago, and the accelerated cosmic rays might have propagated along interstellar magnetic fields connecting the region of Geminga to Earth. Since nothing or very little is known of the local interstellar medium properties, cosmic ray diffusion is not sufficiently constraint to provide a coherent scenario that can explain the observations without considerable fine tuning.

Due to the coincidence of the most significant localized excess observed by Milagro with the heliotail, it is possible that we are seeing the effects of neutron production in the gravitationally focussed tail of the interstellar material, as suggested by~\cite{drury}. Cosmic rays propagating through the direction of the tail interact with matter and magnetic fields to produce neutrons and hence a localized excess of cosmic ray in that direction. But while the target size has about the right size compared to the decay length of multi-TeV neutrons ($\sim$ 0.1 pc), the increase of the gravitating matter density is too low to account for the observed excess.

%It is possible to argue that the large angular scale anisotropy in cosmic rays arrival direction might be generated by a combination of astrophysical phenomena, such as the distribution of nearby recent supernova explosions~\citep{erlykin}, particularly in conjunction with the observed positron anomaly~\citep{yuksel09,blasi} and a possible electron anisotropy as well~\citep{ackermann10}. 

In~\cite{malkov} it is proposed that cosmic rays emitted by a source (like a supernova remnant for instance) within a few 100 pc, are scattered by a strongly anisotropic Alfv\'en wave spectrum formed by the turbulent cascade across the local field direction. Cosmic rays with small pitch angle with respect to the local interstellar magnetic field undergo the highest scattering, thus producing a faint localized excess region. An outer scale of the interstellar medium turbulence of about 1 pc would explain the observations. %Also in this case it is assumed that the existence of two separate localized regions in the northern hemisphere is due to particle trajectory splitting by the magnetic field along the heliotail.

The fractional excess relative to the cosmic ray background observed by Milagro in the direction of the heliotail is $\sim$ 6 $\times$ 10$^{-4}$, i.e. about 1/10 the amplitude of the global anisotropy at TeV energy. This is comparable to the amplitude that the broad tail-in excess would have if extrapolated from 100's GeV to TeV energies~\citep{nagashima}. Such an excess was found to be consistent with hadronic cosmic rays with an energy spectrum of the form $N(E)\,\sim\,E^{-\gamma}\,e^{-E/E_c}$ with spectral index $\gamma\,<\,$2.7 (i.e. flatter than the average cosmic ray spectrum) at 4.6 $\sigma$ level, and a cut-off energy $E_c\,=\,3\,-\,25$ TeV~\citep{abdo2}. A similar spectral hardening was observed by ARGO-YBJ~\citep{argo12}.

In this paper we discuss the scenario where the excess region of cosmic rays from the direction of the heliotail observed from a few tens GeV to about 10 TeV, is generated by re-acceleration processes of a fraction of energetic particles propagating through magnetic reconnection regions along the heliotail. A concurrent contribution from scattering with the turbulence ripples of the heliospheric magnetic field cannot be excluded, especially in relation to the other observed localized fractional excess regions, although this possibility is the topic of another paper~\citep{locscat}.

\begin{figure}[!t]
\includegraphics[width=\columnwidth]{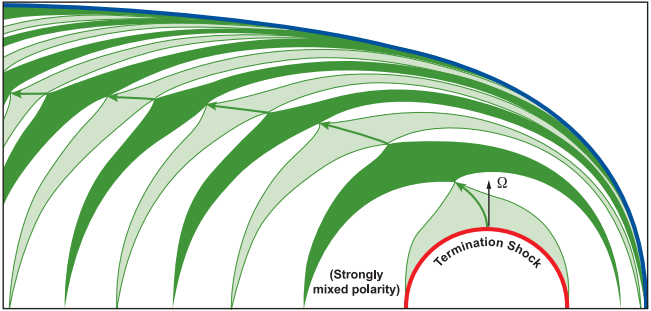}
\caption{A schematic meridional view of the sectored heliospheric magnetic field arising from the 11-year solar cycle polarity reversals. The subsonic solar wind pushes the sectors across the termination shock along the heliotail, compressing them to each other and toward the heliopause. The turbulence that is expected to perturb the heliospheric magnetized plasma, determines the thickness of the current sheet outflow regions. While their length depends on the average magnetic field geometry and by the turbulence level. Adapted from~\cite{nerney, lazopher}.}
\label{structure1}
\end{figure}

%**********************************************************************************************************************************
\section{Magnetic field structure at the heliotail}
\label{sec:magfi}

The motion of solar system through the local partially ionized medium\footnote{the solar system is located at the edge of the so-called local interstellar cloud, which is part of a complex cloudlet structure expanding from the Scorpion-Centaurus Association (see~\cite{frisch2011a})} produces a comet-like interface due to the solar wind plasma advected downstream by the interstellar flow, called the heliosphere. A termination shock, where the solar wind pressure equals that from the interstellar flow, is formed at approximately 100 AU from the Sun. The interface separating interplanetary and interstellar magnetic fields, called heliopause, is at a distance of approximately 200 AU in the upstream direction, and it may extend downstream several thousands AU~\citep{izmodenov} where it could be about 600 AU wide~\citep{pogorelov}. The LIMF drapes around the heliosphere imprinting a deformation that affects its internal structure as well~\citep{pogorelovb}. The heliospheric magnetic field has been studied with detailed MHD simulations, where the effects from the 26 day solar rotation and the 11 year solar cycle were considered~\citep{pogorelov} (see also ~\cite{scherer03}). %The heliospheric plasma flow deviates magnetic field lines toward the tail, where the subsonic solar wind speed is of order 100 km/sec and decreasing until the tail dissolves into the interstellar medium.
Over solar cycles the magnetic field polarity is reversed every 11 years, generating unipolar regions dragged along the heliotail by the $\sim$ 100 km/s solar wind~\citep{par79}. In particular these magnetic regions grow to their maximum latitudinal extent during solar minimum (about 200-300 AU in size) and reduce to zero at solar maximum, when the heliospheric plasma is dominated by the strongly mixed polarity domains (about 0.1-1 AU in size) from solar rotation~\citep{nerney}. Due to the tilt of the solar magnetic axis with respect to its rotation axis, the unipolar regions are thinner at lower latitudes (as shown in Fig.~\ref{structure1}). Therefore the tailward line of view is dominated by the finely alternating magnetic field, while along sighlines away from it the magnetic domains have larger size. MHD numerical simulations show that the sectored unipolar magnetic field regions can propagate for several solar cycles before they dissipate into the local interstellar medium. The corresponding periodic variations on the heliospheric plasma induce changes in the magnitude of the Alfv\'en velocity by about 20\%, and of the magnetic field by about 25\%~\citep{pogorelov}.

There is observational evidence that the plasma in the heliosheath has Reynolds number $Re \approx$ 10$^{14}$ (see~\cite{lazopher} and references therein), meaning that the strength of non-linear convective processes at the largest scale is more important than the damping viscous processes in the dynamics of the flow. We expect a similarly high Reynolds number in the inner heliotail as well. In such conditions it is very unlikely that plasma flow stays laminar, and the downstream motion in the heliotail is likely turbulent. In addition, the presence of neutral atoms in the partially ionized local cloud medium (where the solar system is moving) is essential for the dynamics of the heliosphere and LIMF interaction. Charge-exchange processes between the interstellar inflowing neutral atoms and the outflowing solar wind protons can produce Rayleigh-Taylor type instabilities on the heliopause with amplitude of a few tens AU and over a time scale of a few hundreds years~\citep{liewer96}. Also in a model of plasma-neutral fluid coupled via collision and charge-exchange processes, it is found that such non-linear coupling leads to alternate growing and damping of Alfv\'enic, fast and slow turbulence modes, at $L \sim$100's AU scale and with evolution time longer than inertial time $L/V_A$~\citep{zank}, with $V_A$ the Alfv\'en velocity. Such modulations can propagate on the heliopause, producing ripples along the heliotail that can penetrate deep inside the heliosheath and propagate outward into the local interstellar medium. Therefore, although more investigations are needed in order to understand the detailed plasma properties in the heliotail and its outer boundary, it is reasonable to assume here that magnetic fields in the heliotail are weakly stochastic, and likely reconnecting as the gradient in solar wind advects magnetic field lines closer to each other. The Alfv\'en velocity of the turbulence in the heliotail is expected to be approximately 40-70 km/s, with the actual value depending on the location within the sectored magnetic field~\citep{pogorelov}. This is smaller than the solar wind speed downstream the termination shock, therefore magnetic reconnection in the heliotail is not expected to change the overall magnetic field structure. Nevertheless, the effects of turbulence are very important from the point of view of magnetic reconnection and the particle acceleration that it entails.

Simulations of the magnetic fields in the heliotail are extremely challenging due to its extension and to the complex interaction with the interstellar wind and between heliospheric magnetic field and the LIMF, but mainly because there is currently no direct data collection from this remote portion of the heliosphere. Future refinements of MHD simulations will provide higher resolution mapping of the heliotail and of the plasma properties, that will help improving our knowledge of its effects on TeV cosmic ray propagation.

%**********************************************************************************************************************************
\begin{figure}[!t]
\center{\includegraphics[width=0.6\columnwidth]{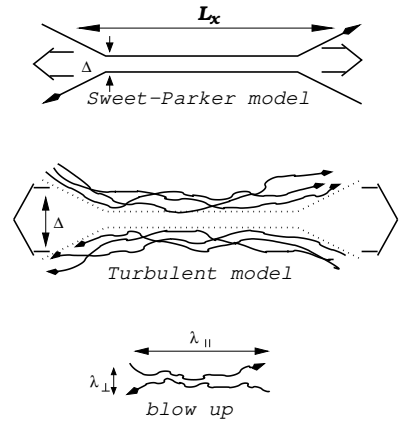}}
\caption{{\it Upper plot}: 
Sweet-Parker model of reconnection~\citep{sweet58, parker57}. The outflow region is limited within a thin transition zone $\Delta$ between the reversed magnetic field lines, which depends on plasma resistivity. The other scale is an astrophysical scale $L\gg \Delta$. {\it Middle plot}: Reconnection of weakly stochastic magnetic field according to~\cite{lv99}. The outflow region is determined by the diffusion of magnetic field lines, which depends on the field stochasticity. {\it Lower plot}: An individual small scale reconnection region. The reconnection over small patches of magnetic field determines the local reconnection speed. The global reconnection speed is substantially larger as many independent patches come together. The bottleneck for the process is given by magnetic field wandering and it gets comparable to $L$ as the turbulence injection velocity approaches the Alfvenic one. From~\cite{laz04}.}
\label{fig:recon1}
\end{figure}

%**********************************************************************************************************************************
\section{Stochastic magnetic reconnection}
\label{sec:rec}

\begin{figure*}[!t]
\center{
\includegraphics[width=\columnwidth]{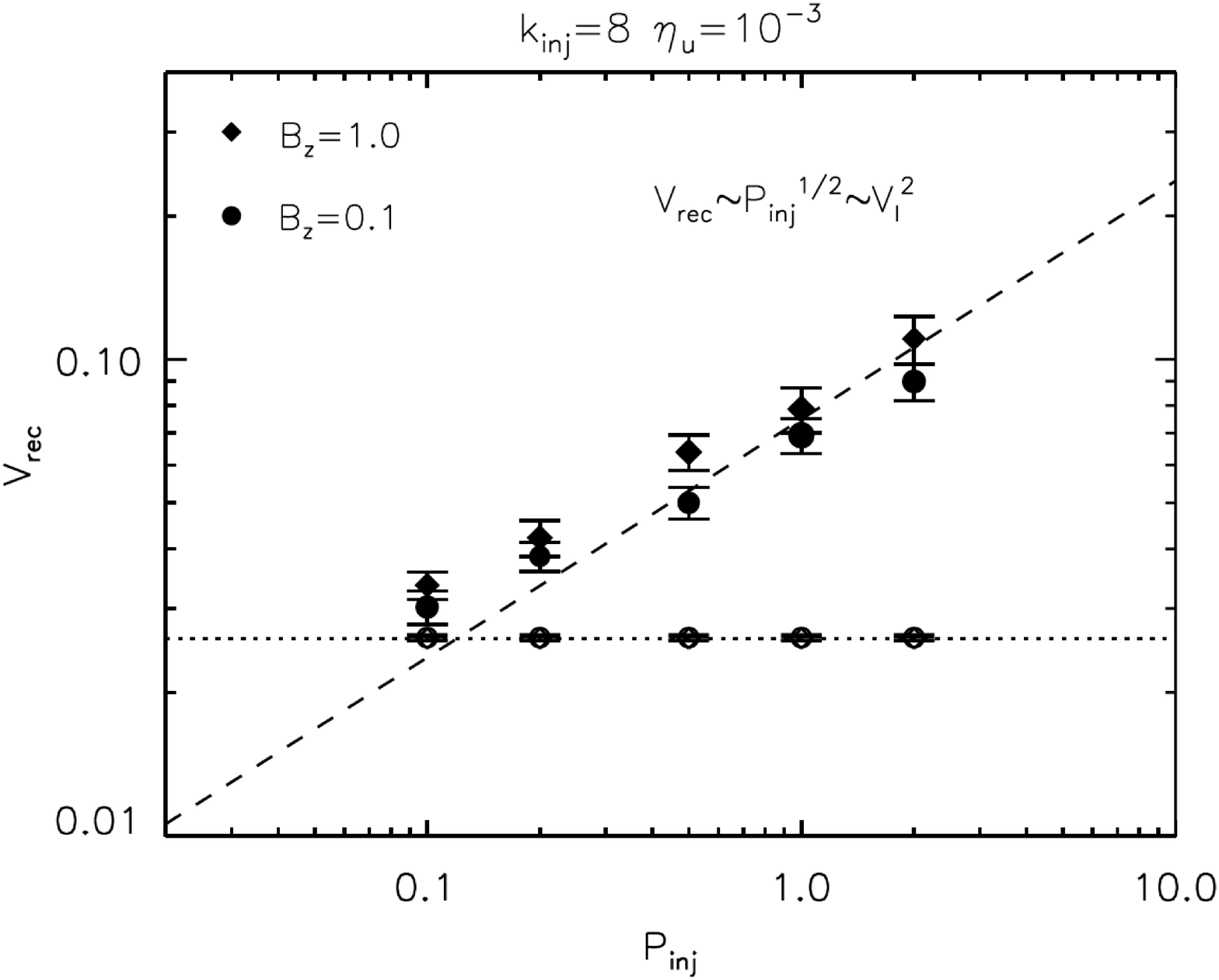}
\includegraphics[width=\columnwidth]{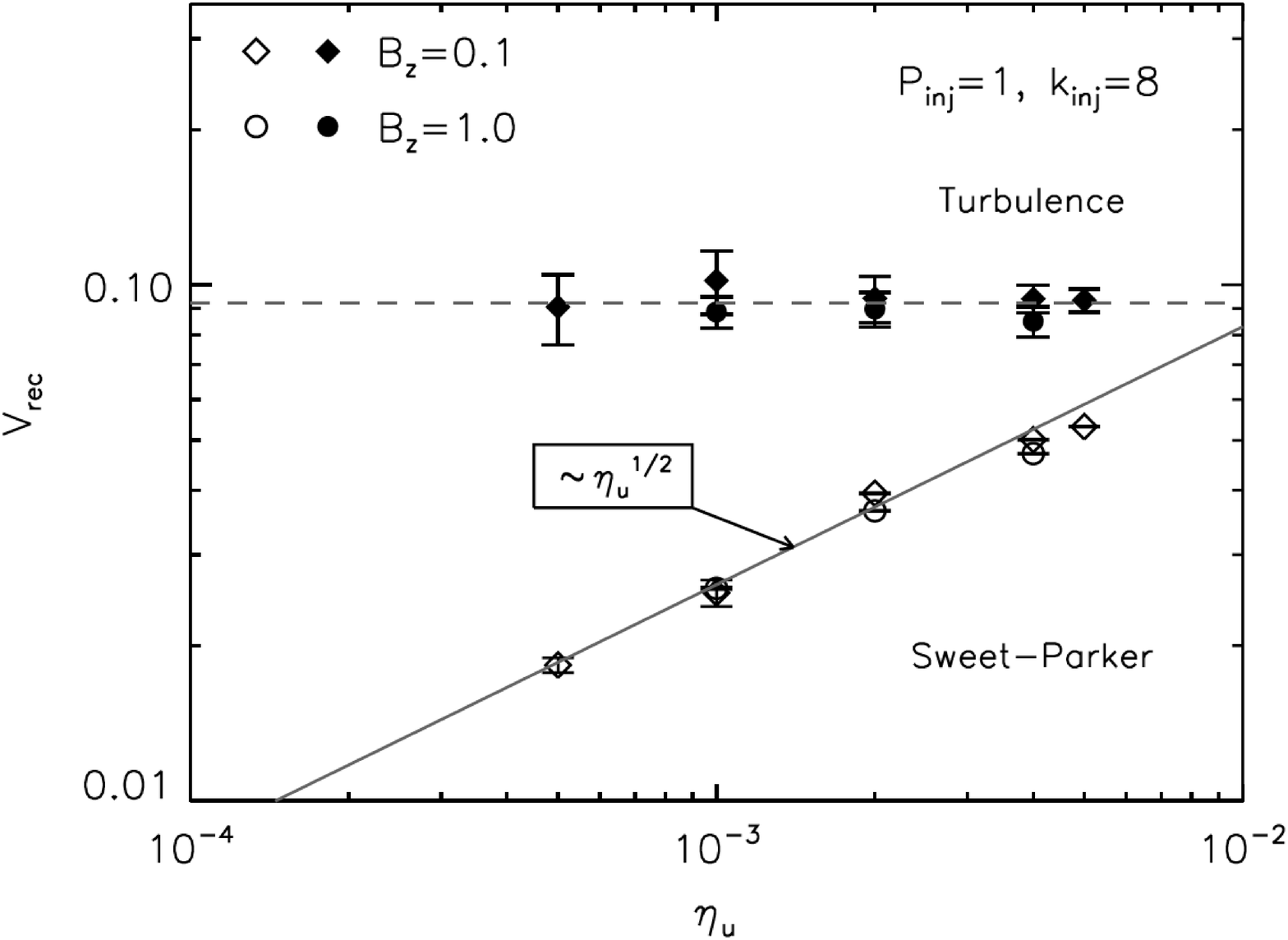}
}
\caption{{\it Left panel}: dependence of the reconnection speed $V_R$ on injection power $P_{inj}$.  {\it Right panel}: dependence of the reconnection speed $V_R$ on the uniform resistivity $\eta_u$. Open symbols are for Sweet-Parker reconnection scenario~\citep{sweet58, parker57}, and filled symbols are for weakly stochastic reconnection scenario~\citep{lv99}. See~\cite{kowal,laz11}.}
\label{fig:recon2}
\end{figure*}

Astrophysical plasmas are often highly ionized and magnetized~\citep{par70}, and they undergo dissipative processes, which annihilate the magnetic fields and convert electromagnetic energy into plasma energy. Due to these processes, plasma from regions of a given polarity becomes magnetically connected to that of opposite polarity: this is when magnetic reconnection occurs. However, reconnection speed, and therefore the rate at which magnetic energy is converted into plasma energy, is too small to be important for acceleration of energetic particles, unless the effects of plasma resistivity are negligible.

In the Sweet-Parker model of reconnection~\citep{sweet58, parker57} the outflow is limited within the transition zone $\Delta$, which is determined by Ohmic diffusivity (see top of Fig.~\ref{fig:recon1}). In this model reconnection speed is smaller than the Alfv\'en velocity of the plasma by a factor equal to $S^{-1/2}\,=\,(L\,V_A/\eta)^{-1/2}$, where $S$ is the Lundquist number, $L$ the length of the current sheet, $V_A$ the Alfv\'en speed and $\eta$ is the Ohmic resistivity of the plasma. The length of the current sheet is determined by the extent of magnetic flux tubes that get in contact. Although the properties and dimensions of the heliotail are not well constraint yet, it is possible to state that the extension of current sheets between sectored heliospheric magnetic field in the heliotail could lay between about 100 AU and 300 AU~\citep{pogorelov}. Assuming the same plasma properties as in the heliosheath closer to the heliospheric nose, S is about 10$^{12}$-10$^{13}$~\citep{lazopher}. Therefore, the corresponding reconnection speed for the Sweet-Parker model is several orders of magnitude smaller than the Alfv\'en velocity. In fact in this case plasma collected over the size L should be ejected with speed $\sim \,V_A$ through the outflow region of thickness $\Delta = L\,S^{-1/2}$, i.e. much smaller than the length of the current sheet. It is the large difference between $L$ and $\Delta$ that makes reconnection slow and unlikely to produce any effect on the plasma. The major consequence of such a model is that reconnection speed is limited by Ohmic resistivity of the plasma (see right panel of Fig.~\ref{fig:recon2}). Since most astrophysical plasmas have very low resistivity, a magnetic reconnection mechanism such this would not produce observable effects.

On the other hand, various observations suggest that reconnection, when it occurs, can be fast in some circumstances. For instance the development of solar flares suggests that magnetic reconnection should be initially slow in order to ensure the accumulation of magnetic flux, and then suddenly becomes fast in order to explain the observed fast release of energy. Fast reconnection would require $L\,\sim\,\Delta$, meaning that the region over which magnetic flux tubes intersect is comparable to the size of the outflow region. This can be achieved by increasing the outflow region beyond the prediction of Sweet-Parker model, or by making $L$ as small as the Ohmic diffusion region, so that magnetic field lines reconnect in an "X-point". In this X-point collisionless model~\citep{petschek64}, reconnection speed does not depend on the resistivity and it is of the order of Alfv\'en velocity of the plasma. On the other hand X-points are found to be unstable and to collapse into a Sweet-Parker current sheet in the MHD regime~\citep{biskamp96}. In a collisionless plasma X-point, stability can be maintained through coupling to a dispersive plasma mode~\citep{sturrock66,shay98b}. Recently it was discovered that X-points can be stabilized in the presence of MHD Hall effect so that the outflow opens up on larger scale, thus making reconnection fast~\citep{shay98,shay04}. On the other hand, most astrophysical plasmas are turbulent, and the heliosphere is most probably not an exception. This means that X-points can be created by turbulence and sustained at small scale by Hall-MHD effects, but only until turbulence itself collapses the X-points to form extended thick outflow regions (as observed by~\cite{ciaravella} in multi-frequency observations of solar flares).
%(see for instance~\cite{lazopher} where a model of acceleration of anomalous cosmic rays by reconnection in the heliosheath is proposed).
Even without turbulence it has been suggested that magnetic islands dynamically produced at X-points tend to be volume filling and to produce thick collisionless reconnection regions~\citep{drake} with high reconnection speed. However, since turbulence is likely ubiquitous in astrophysics plasmas, we concentrate here on fast reconnection mechanisms in weakly stochastic plasmas.

A model of fast magnetic reconnection that generalizes the Sweet-Parker scheme for the case of weakly stochastic magnetic fields was proposed by~\cite{lv99} (henceforth LV99). Even though the notion of reconnection affected by turbulence is not new, in the LV99 model it is recognized that turbulence can decouple the width of plasma outflow region from the scale determined by Ohmic effects. In fact the outflow width is limited by the diffusion of magnetic field lines, which depends on turbulence only (see center of Fig.~\ref{fig:recon1}), and can be much wider than the thickness of the individual current sheets (see bottom of Fig.~\ref{fig:recon1}). Although reconnection events happen on small scales $\lambda_{\parallel}$, where magnetic field lines get into contact, a number of independent reconnection processes takes place simultaneously over extended macroscopic current sheets $L\,\gg\,\lambda_{\parallel}$ within a wide outflow region $\Delta \sim L$. Therefore the effective reconnection rate is not limited by the speed of individual Sweet-Parker events on scale $\lambda_{\parallel}$ (where plasma resistivity plays a dominant role), instead, it is enhanced by the large $\Delta$, that depends on the magnetic field wandering. %Reconnection speed, as a consequence, is large and comparable to the Alfv\'en velocity of the plasma.
%The LV99 model of reconnection considers 3D configurations of stochastic magnetic field lines that can enter the reconnection region and reconnect there independently.
In such a situation it was found that reconnection speed is close to the turbulent velocity in the plasma. In particular, assuming isotropically driven turbulence characterized by an injection scale $l \lesssim L$ the reconnection speed is~\citep{lv99,laz06}

\begin{equation}
V_R\,\approx \,V_A\,\left(\frac{l}{L}\right)^{1/2}\,\left(\frac{V_l}{V_A}\right)^2,
\end{equation}
where $V_l$ is the turbulent velocity at the largest scale and $V_A$ the Alfv\'en velocity. Since turbulence in the heliotail is assumed to be weak, magnetic perturbations are comparably smaller with respect to the mean heliospheric magnetic field, therefore $V_l \lesssim V_A$, i.e. turbulence in the heliotail is sub-Alfv\'enic.
%If $V_{turb}\,=\,V_A\,(V_l/V_A)^2$ is the velocity at the scale where turbulence transits from weak to strong MHD turbulence (see~\cite{lv99,laz06}), then reconnection speed can be expressed as
%\begin{equation}
%V_R\,\approx \,V_{turb}\,\left(\frac{l}{L}\right)^{1/2}.
%\label{eq:rec}
%\end{equation}

Numerical MHD calculations of weakly stochastic magnetic reconnection were performed by~\cite{kowal} and they proved that reconnection is fast and independent of Ohmic resistivity of the plasma, as shown in right panel of Fig~\ref{fig:recon2}, while it depends on the power of injected turbulence (shown in left panel of Fig.~\ref{fig:recon2}). In these simulations, turbulence is preexisting and not related to reconnection processes. By varying the Ohmic and anomalous resistivity of the plasma, reconnection rate is not affected, confirming that in the presence of turbulence resistivity is not important.
%In fact in the case of weak turbulence the relation between the power and the injection velocities are different from the usual Kolmogorov estimate, namely $P_{inj}\,\sim \,V_l^4\,/(l\,V_A)$~\citep{lv99}. Therefore reconnection speed can be expressed as
%\begin{equation}
%V_R\,\approx \,\left(\frac{P_{inj}}{L\,V_A}\right)^{1/2}\,l.
%\end{equation}

In the LV99 model, reconnection develops while the wide outflow region is filled with turbulent reconnected magnetic field lines moving in opposite directions. In fact, numerical simulation by~\cite{kowal11} show that the interface between the oppositely directly magnetic fields has a much more complex topology if compared to the Sweet-Parker reconnection mechanism, and also to the schematic representation in Fig.~\ref{fig:recon1} (see Sec.~\ref{sec:acc}). The outflow volume is filled with enhanced current density regions with loops of reconnected magnetic flux, where locally reconnection works faster since the current density reaches higher values. The magnetic loops shrink as a dynamical consequence of reconnection development, while multiple reconnection events happen at the same time due to the magnetic field stochasticity.

%Fig.~\ref{fig:recon2} shows the dependence of reconnection speed on the turbulence injection power and on the plasma uniform resistivity, as obtained from numerical calculations~\citep{kowal}. While in the Sweet-Parker scenario, the reconnection speed depends on the resistivity of the plasma and it does not depend on the turbulence power, in the Lazarian \& Vishniac scenario it shows no dependency on resistivity and it increases with injected power of the turbulence, as predicted in~\cite{lv99}. The fast nature of the weakly stochastic magnetic reconnection mechanism is a consequence of the turbulence in the plasma. Astrophysical plasmas are by nature turbulent and likely have low resistivity, therefore stochastic magnetic reconnection processes provide an efficient mechanism to transfer electromagnetic energy into plasma energy.

%**********************************************************************************************************************************
\section{Acceleration in reconnection regions}
\label{sec:acc}

Electric fields associated with reconnection events can accelerate energetic particles. For a particle of charge $q$, the typical energy gained in a reconnection process is of the order $q\,(V_R/c)\,B\,\lambda_q$, where $\lambda_q$ is the coherence length of the particle within the reconnection layer. Efficient acceleration would require, therefore, both $V_R$ and $\lambda_q$ to be large. However, in general in any fast reconnection mechanism, the fraction of volume that is subject to resistive effects and reveals strong electric fields is small and most of the magnetic energy is converted into kinetic energy of the plasma instead. Therefore only a small fraction of the energy can be transferred through any fast reconnection process to energetic particles if direct electric field is involved. The observation of a large normal component of the electric field near an X reconnection point in the Earth's magnetotail, was interpreted as Hall electric field at the X-point current sheet by~\cite{wygant05}, capable of accelerating ions to $\sim$ 10's keV scale. In this paper we concentrate on the mechanism of energetic particle re-acceleration in weakly stochastic reconnection regions.

\begin{figure}[!t]
\begin{center}
\center{\includegraphics[width=0.6\columnwidth]{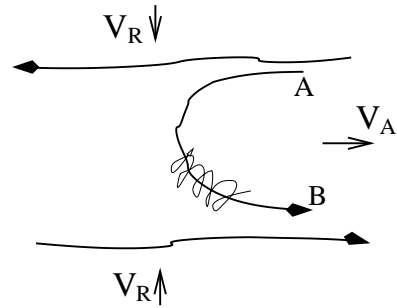}}
\caption{
The simplest realization of magnetic reconnection at small scale and of acceleration as an energetic particle bounces back and forth between converging magnetic field lines. The converging velocity determines the reconnection speed $V_R$, while the advection of the accelerated particles entrained on the magnetic field lines, occurs at an outflow speed that in most cases is of order the Alfv\'en velocity of the plasma $V_A$. Particles bouncing at points A and B happens because either of streaming instability induced by energetic particles or magnetic turbulence in the reconnection region. In an actual turbulent plasma, the outflow region, at large scale, is filled with reconnecting loops and current sheets, each of which a possible acceleration site (see text). From~\cite{laz05}.}
\label{fig:recon3}
\end{center}
\end{figure}

In the LV99 mechanism, reconnection speed can approach $V_A$, which can be appreciably large, therefore particles entrained on reconnecting field lines bounce back and forth between the approaching magnetic walls while staying on the field lines that are contracting. This results in an increase of particle velocity with every bouncing, as discussed by~\cite{GL03,gouveia} (see also~\cite{laz05}) where it was shown that reconnection induces particle acceleration. The effect that individual magnetic loops shrinking in the reconnection region have on energetic particles, is equivalent to that of first order Fermi acceleration in magnetic mirrors. Fig.~\ref{fig:recon3} schematically represents the simplest realization of acceleration within the reconnection region expected within LV99 model. As energetic particles bounce back and forth between converging magnetic fluxes, they gain energy. In the figure, the bouncing at points A and B is just an illustration of the process. In reality particles never pass by the same points in 3D, but they locally stream along magnetic field lines and bounce back and forth through magnetic bottles that form in the reconnection region.

The simple acceleration process represented in Fig.~\ref{fig:recon3} can be easily quantified. An energetic particle with energy $E$ bouncing back and forth between a magnetic mirror will gain an energy $\Delta E\,\sim \, (V_R/c)\, E$ in every cycle. The process continues until particles gain enough high energy to either diffuse perpendicularly out of the reconnection region or get ejected by the outflow plasma at the Alfv\'en velocity. This last possibility was considered by~\cite{GL03,gouveia}, namely that particle diffusion perpendicular to the mean magnetic field is negligible. %has speed $V_{\perp, diff}\,\ll \, V_R$ and that the parallel diffusion has speed $V_{\parallel, diff}$ does not exceeds $V_R$.
Perpendicular diffusion arises from magnetic field wandering as particles scatter marginally perpendicular to the local magnetic field. This effect was accounted by~\cite{yan04,yan08} to describe cosmic ray propagation, for instance. %For the developed strong turbulence at scales $\Delta$ less than turbulence injection scale $l$, the perpendicular diffusion speed for energetic particles can be expressed as $V_{\perp, diff}\,\approx \, V_{\parallel, diff}\,\left(\frac{V_R}{V_A}\right)^2\,\left(\frac{l}{L}\right)$~\citep{yan08}, which means that for sub-Alfv\'enic turbulence with $l\,\sim \,L$ the constraint on perpendicular diffusion velocity is less restrictive than $V_{\parallel, diff}\,<\,V_A$.
As mentioned, the properties of turbulent plasma in the heliotail are not well know at this point, therefore it is difficult to quantify the diffusion regime, nevertheless, in general perpendicular scattering in sub-Alfv\'enic turbulence is found to be subdominant with respect to parallel scattering~\citep{yan08,beres11} (see also~\cite{laz06,laz06err}). In this case, the energy spectrum of accelerated test particles\footnote{i.e. neglecting the back-reaction of accelerated particles, see~\cite{longair}} is~\citep{GL03,lazopher}

%it is possible to derive the energy spectrum of particles undergoing first order Fermi acceleration in weakly stochastic reconnection regions assuming a test particle approximation, i.e. neglecting the back-reaction of the accelerated particle (see~\cite{longair}). Considering the acceleration of $N_0$ particles with initial mono-chromatic energy $E_0$, in a process cycle involving the bouncing of a particle from one end to the other of the shrinking reconnection region and back. After each cycle the particle energy is $E\,=\,\beta\,E_0$, therefore after $n$ cycles, the final energy is $E_n\,=\,\beta^n\,E_0$. At the same time if $P$ is the probability that after each cycle the particle remains within the acceleration region, after $n$ cycles the number of confined particles is $N_n\,=\,P^n\,N_0$. This means that the number of particles accelerated to an energy of at least $E$ after a number of cycles is
%\begin{equation}
%N(\geq E)\,=\,N_0\,\left(\frac{E}{E_0}\right)^{lnP/ln\beta},
%\end{equation}
%therefore the differential spectrum arising from the acceleration mechanism is
%\begin{equation}
%N(E)dE\,\sim \,E^{-1\,+\,(lnP/ln\beta)}\,dE,
%\end{equation}
%where $\beta \,=\,1\,+\,\frac{8}{3}\,\left(\frac{V_R}{c}\right)$ and $P\,=\,1\,-\,4\,\left(\frac{V_R}{c}\right)$ in case $V_{\perp, diff}\,\ll \, V_R$, i.e. assuming that particles escape by advecting out the reconnection region with the magnetized plasma outflow (see~\cite{GL03,lazopher}). Since $V_R/c\,\ll \,1$ the spectrum arising from acceleration in a single reconnection region is
\begin{equation}
N(E)dE\,\sim \,E^{-5/2}\,dE.
\label{eq:spectrum}
\end{equation}
If perpendicular diffusion in the reconnection region is not negligible, particles bouncing between approaching reconnecting field lines of the magnetic mirror are not confined as within walls, but can keep bouncing while reconnection proceeds. In this situation particles may have cross field propagation but cannot escape from the large scale reconnection region, producing a spectrum asymptotically reaching $N(E)dE\,\sim \,E^{-1}\,dE$.

\begin{figure}[!t]
\begin{center}
\center{\includegraphics[width=\columnwidth]{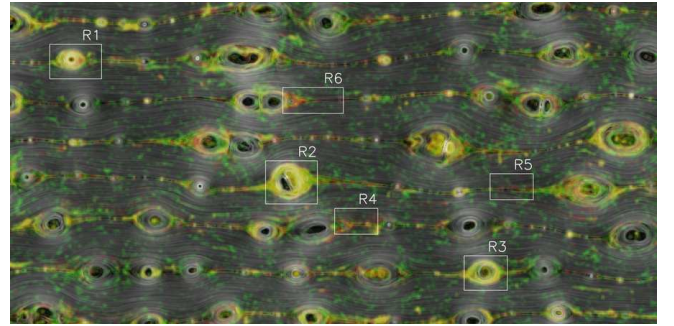}}
\caption{Evolved 2D snapshot of magnetic field configuration where eight parallel Harris current sheets were perturbed in order to trigger plasma instabilities and reconnection events (represented in grey shades). 10,000 test particles, with initial thermal distribution with temperature corresponding to the sound speed of the MHD model, were injected in this plasma snapshot to study the acceleration mechanism induced by magnetic reconnection. The red and green colors correspond to regions where either parallel or perpendicular acceleration occurs, respectively, while the yellow color shows locations where both types of acceleration occur. The parallel component increases in the contracting islands and in the current sheets as well, while the perpendicular component increases mostly in the regions between current sheets. The white boxes correspond to the sites where detailed determination of acceleration properties were done (see text). From~\cite{kowal11}.}
\label{fig:topology}
\end{center}
\end{figure}

In case of re-acceleration of cosmic rays with a seed spectrum $E^{-2.7}$, after acceleration it still becomes $\propto E^{-5/2}$, i.e. harder than the initial spectrum. It is important to note that the expected energy spectrum Eq.~\ref{eq:spectrum} is an estimate based on a rather idealized situation. The derivation above considers only particles bouncing back and forth between the two reconnection layers. The actual picture of stochastic reconnection in the LV99 model includes many simultaneous reconnection events happening at different scales throughout the reconnection volume. Fig.~\ref{fig:topology} (from~\cite{kowal11}) shows an evolved 2D snapshot of magnetic field configuration during reconnection from a nearly incompressible non-resistive MHD domain simulation without including kinetic effects, such as pressure anisotropy, the Hall term, or anomalous effects. The initial condition of the domain was set with eight Harris current sheets in a periodic box and a density profile corresponding to a uniform total pressure. A perturbation with random weak velocity fluctuations was used to enable spontaneous reconnection events. It is evident from the figure that the reconnection volume is filled with magnetic loops (or islands) and that several reconnection events occur at the same times within the loops and along the current sheets between the loops. The simulation shows the existence of merging loops with their resulting deformation and contraction, that provide appropriate conditions for particle acceleration. This picture is very similar to the 2D simulation by~\cite{acr2}, where islands, or loops, are in fact only the 2D projections of 3D magnetic tubes, as shown in~\cite{kowal11}

The topological complexity of the reconnection region may have an influence on the actual spectral shape of the accelerated particles. Each local reconnection region (within a magnetic loop or a current sheet) provides the accelerated spectrum $\propto E^{-5/2}$. But when energetic particles cross several reconnection regions they undergo further acceleration with a seed spectrum corresponding to that gained within the previously crossed reconnection region. It is well known from the theory of diffusive shock acceleration, that the supply of a seed power law spectrum into an acceleration region leads to an amplification of the distribution without changing the spectral index. However the increased number of high energy particles is accompanied by a decrease of the number density at the low energy cutoff, leading to a flattening of the distribution at intermediate energy (see~\cite{bell,melrose,gieseler}), and, therefore, to a harder energy spectrum. The exact slope of the final spectrum after bouncing within and escaping from several reconnection regions depends on the cutoff energy and on energy loss processes that particles undergo during acceleration and, most importantly, between acceleration processes. Scattering, for instance can degrade particle energy so that at each acceleration step a new low energy population is seeded into the process, leading to a softening of the spectrum.

Non-linear effects from back-reaction of the accelerated particles might be important in reconnection processes. However so far the only approach to address back-reaction of particles on reconnecting plasma involved electrons~\citep{drake}. In fact the evidence of back-reaction can be found in the simulations of test particles propagating in the magnetotail~\citep{birn} and also in test particles studies in MHD models with magnetic islands~\citep{matthaeus,kliem}. In~\cite{drake}, where acceleration occurs in contracting loops formed in 2D collision-less reconnection, back-reaction is introduced by the term $(1\,-\,8\pi \bar{\epsilon_{\parallel}}/B^2)$, where $\bar{\epsilon_{\parallel}}$ is the energetic particle parallel energy averaged over the distribution of particle velocities. This would produce an accelerated spectrum $\sim \,E^{-3/2}$, instead of the steeper $\sim \,E^{-5/2}$.

%Since the plasma properties in the heliotail are not well known it is difficult at this point to quantify particle diffusion and scattering processes, which at the moment can be assumed to be similar to those measured in more explored portion of the heliotail. On the other hand, if a fraction of cosmic rays propagating through the heliotail are accelerated in weakly stochastic magnetic reconnection regions, their accelerated spectrum can be flatter than E$^{-2.7}$, depending on the plasma properties and on the scattering of energetic particles with the turbulence ripples, in general agreement with Milagro observation~\citep{abdo2}.

%\subsection{Anisotropic Acceleration}
%\label{ssec:aa}

\begin{figure}[!t]
\begin{center}
\center{\includegraphics[width=\columnwidth]{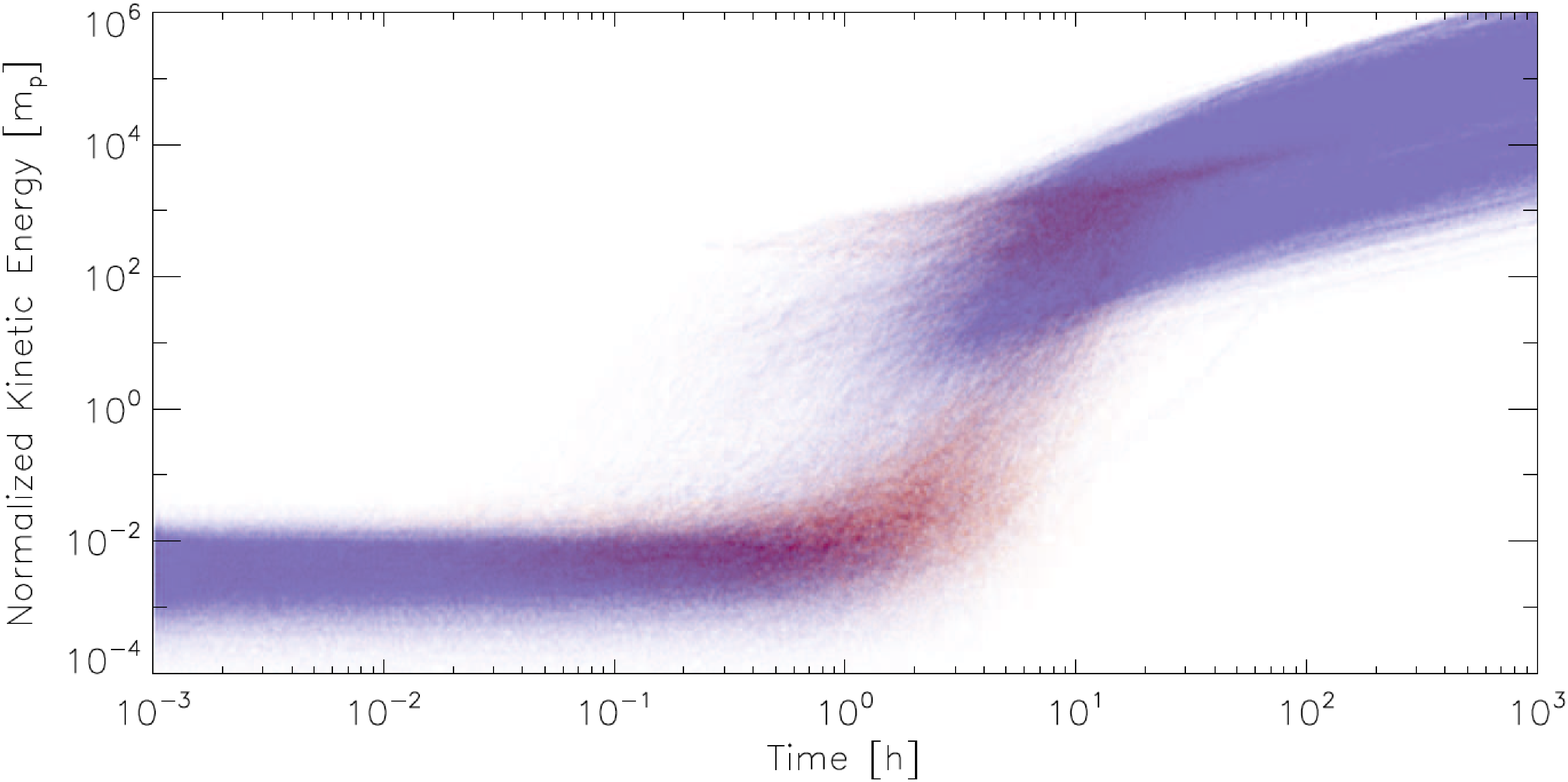}}
\center{\includegraphics[width=\columnwidth]{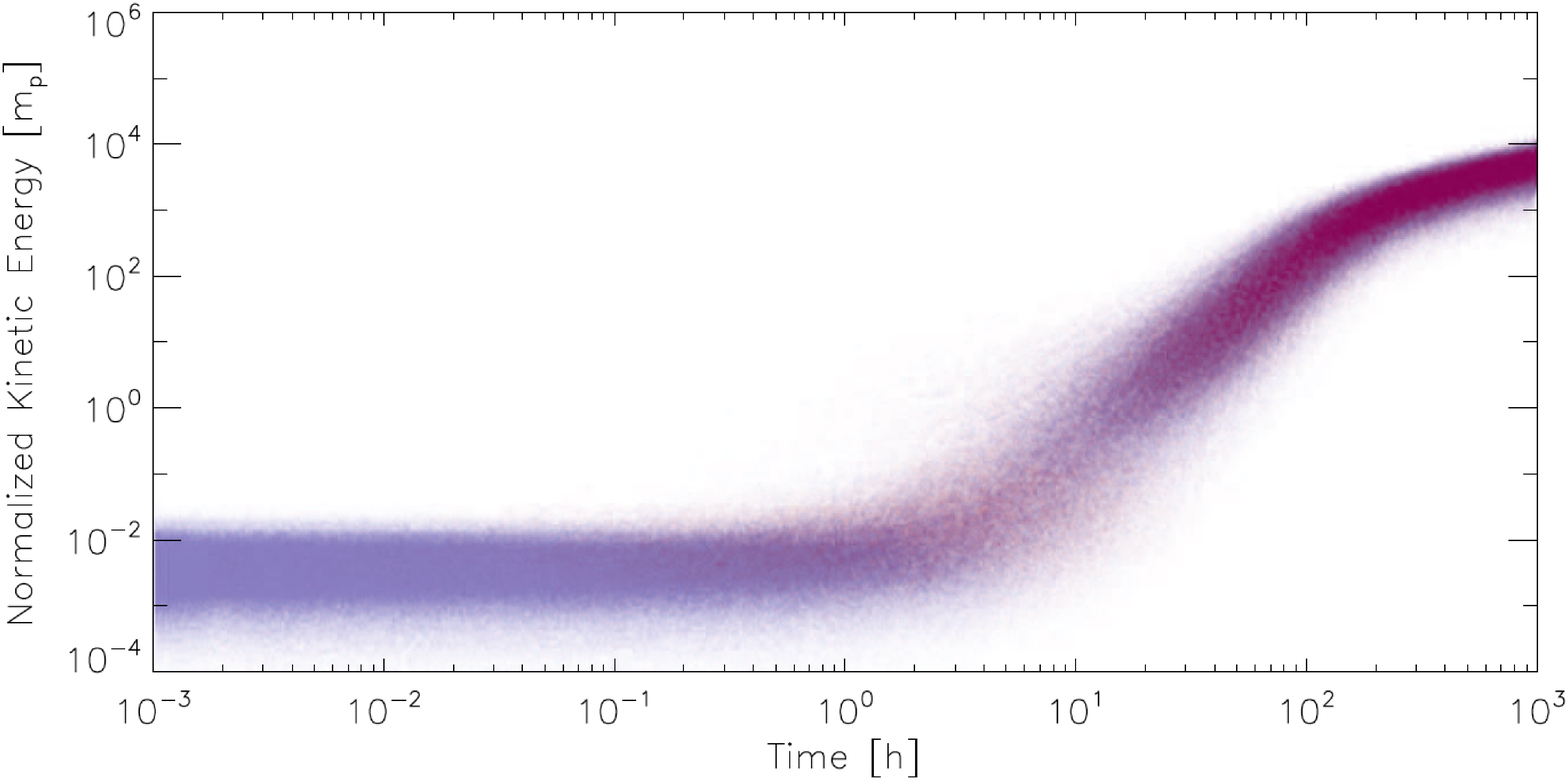}}
\caption{{\it Top panel}: kinetic energy evolution in time of 10,000 protons in a 2D model of reconnection (i.e. with magnetic field on a plane). {\it Bottom panel}: the same distribution but in a fully 3D model of reconnection. In 2D the perpendicular component of particle velocities (in blue) becomes dominant over time with respect to the parallel component (in red). While in a 3D domain it is the parallel component to dominate. The energy is normalized to the proton rest mass. The background magnetized flow with multiple current sheet layers is at time 4.0 in Alfv\'en time units in the model. Note that the transition from exponential energy grow to nearly linear occurs when the largest loop in the plasma reaches the size of a few tens the size of simulation box. From~\cite{kowal11}.}
\label{fig:acceleration}
\end{center}
\end{figure}

Numerical simulations of test particles injected in the domain represented in Fig.~\ref{fig:topology}, with initial thermal distribution, show that particle velocity parallel (in red) and perpendicular (in green) to the mean magnetic field increases. Yellow color indicates the locations where acceleration increases both components without preference. But while in 2D domains perpendicular velocity mostly increases over longer integration time, in 3D, where the loops develop in magnetic tubes, there is no such a limitation making acceleration in the parallel dimension dominant~\citep{kowal11,kowal12} as shown in Fig.~\ref{fig:acceleration}, where velocities were sampled within the regions indicated with white boxes in Fig.~\ref{fig:topology}. It is found that within contracting/deforming magnetic loops or current sheets particles accelerate mainly through first order Fermi acceleration with particles bouncing back and forth between converging mirrors~\citep{GL03,gouveia,acr2}, while outside these regions particles mostly undergo drift acceleration from magnetic field gradients. It is possible that turbulence far from loops, current sheets and diffusion regions, favors mostly second order Fermi acceleration mechanisms with particles being scattered by approaching and receding magnetic irregularities. In~\cite{kowal12} it was argued that second order Fermi acceleration is the dominant process in purely turbulent plasmas with no converging flow, although its rate is reduced. Moreover, reconnection layers in pure turbulence could be responsible for first-order Fermi acceleration of low energy particles. However, more studies are needed to fully understand the interplay between different acceleration mechanisms in turbulent media.

In shock acceleration, particles gain energy via plasma compression differences between the upstream and downstream regions. Energy gain is described by Parker's transport equation~\citep{par65}, which was derived in the limit of strong scattering, and it is explicitly driven by plasma compression. On the other hand in magnetic reconnection it is possible to have acceleration even in an incompressible plasma. In fact numerical MHD simulations such as the one by~\cite{kowal,kowal11,kowal12} were done in a nearly incompressible regime, and as long as there is no strong scattering to maintain plasma isotropy the parallel energy gain dominates and, as a result, the particles entrained in the reconnecting magnetic flux gain net energy.

\subsection{Re-acceleration of Cosmic Rays}
\label{ssec:cr}

The sectored heliospheric magnetic field regions in the heliotail generated by the 11 year solar cycle is composed by 100-300 AU wide unipolar domains with turbulence scale likely of order 10-100 AU, although the injection scale is not known precisely. The downstream solar wind motion in the heliotail induces converging flows in the turbulent plasma that ignite reconnection. As discussed in Sec.~\ref{sec:rec}, turbulence creates the conditions for forming multiple simultaneously reconnecting magnetic fluxes (loops and current sheets) throughout large portions of the plasma. Acceleration takes place across the entire reconnection region and energetic particles are accelerated through a sequence of multiple reconnection events (see Sec.~\ref{sec:acc}). Although second order Fermi acceleration from pure turbulence may occur as well, as long as reconnection is efficient first order Fermi acceleration is dominant. The overall process, therefore, takes place across regions that are comparable with the size of the unipolar magnetic domains, or even the size of the heliotail itself. The energy spectrum of those re-accelerated particles is, in the simplest case, $\sim \,E^{-\gamma}$, with spectral index $\gamma=5/2$ or smaller as discussed in Sec.~\ref{sec:acc}, which is flatter than the mean cosmic ray spectrum. Such acceleration can occur for as long as the cosmic ray particles are trapped within the reconnection regions. Using the simple argument that the gyroradius should not be larger than the size of the magnetized region $L_{zone}$, the maximum energy for a proton is~\citep{longair}

\begin{equation}
E_{max}\,\approx \,0.5\,\left(\frac{B}{1\,\mu G}\right)\,\left(\frac{L_{zone}}{100\,AU}\right)\, TeV.
\label{eq:emax}
\end{equation}

The magnetic field strength in the heliotail is not known with precision, but we can assume that it is of the order of 1-4 $\mu$G~\citep{pogorelov}. $L_{zone}$ is assumed to be within the range 100-500 AU (the higher bound being approximately the heliotail thickness), therefore the maximum energy that cosmic rays can achieve is approximately between 0.5 - 10 TeV. %As mentioned in Sec.~\ref{sec:interp} this is the energy range where Milagro observed a cut-off on the harder cosmic ray spectrum from the localized fractional excess region~\citep{abdo2}.
This means that the fractional excess region observed in the direction of the heliotail is likely expected to have a harder spectrum than the average cosmic rays up to about 1 - 10 TeV. Above this energy the spectrum transitions back to the steeper $\sim \,E^{-2.7}$. Scattering processes within the heliotail can mitigate the acceleration effects and the related cosmic ray distribution at a given energy. Based on the global magnetic field structure in the heliotail, confirmed in MHD simulations (see Sec.~\ref{sec:magfi}), TeV cosmic particles experience the lowest scattering along the line of sights parallel to the interstellar downstream flow, where magnetic field is strongly mixed at small scale. While away from this direction scattering in the unipolar magnetic domains scrambles particles direction and effectively reduces the overall acceleration efficiency. At sub-GeV energies, the stronger scattering along the heliotail would degrade anisotropy and spectral features. As stated in~\cite{reconnection}, the properties of magnetic reconnections are still under extensive study, and their level of complexity is being subject of debate. At the same time the very little explored tail region of the heliosphere, makes the problem under discussion here even more uncertain. However it is suggestive that the observation of TeV cosmic ray arrival distribution and energy spectrum over small angular regions could be used to probe properties over the most remote regions of the heliosphere.

Although there is no energy spectral determination in the sub-TeV energy range, the significant hardening of the spectrum observed by Milagro~\citep{abdo2} and ARGO-YBJ~\citep{argo12} is indicative of a possible re-acceleration mechanism that involves a fraction of cosmic rays propagating from the direction of the heliotail. While waiting for other experimental results that can confirm a harder than average energy spectrum of cosmic rays within the localized excess region, the energy flux corresponding to the $\sim$ 6 $\times$ 10$^{-4}$ fractional excess from $\sim \,10$ GeV to a few tens of TeV can be estimated to be approximately between $10^{-9}$ and $10^{-8}$ erg cm$^{-2}$ s$^{-1}$, for $\gamma \,=$ 2.7 - 2.0, respectively. The corresponding average power dissipated in the re-acceleration of such energetic particles is approximately between $10^{20}$ and $10^{22}$ erg s$^{-1}$. Even though a precise quantitative assessment of the power necessary to re-accelerate the fraction of energetic cosmic rays forming the Milagro localized excess region, is not possible at this point, this simple estimation shows that the fraction of heliospheric plasma power dissipated into cosmic ray kinetic energy is very small if compared to that transported by the solar wind ($\approx \,10^{27}$ erg s$^{-1}$, see~\cite{parker62}).

It is interesting to note that within the last few years experimental evidence that cosmic ray spectrum becomes harder at about 0.2-0.3 TeV/nucleon has been accumulated by ATIC-2~\citep{atic2}, CREAM~\citep{cream} and PAMELA~\citep{pamela}. In particular the CREAM results seem to suggest that cosmic ray spectrum may become softer again at about 10 TeV/nucleon, although more observation is needed on this regard. The direct observation of a correlation between spectral features and arrival direction would provide a breakthrough on the role of the heliotail in the TeV cosmic ray properties.

%**********************************************************************************************************************************
\conclusions
\label{sec:conc}

The observation that cosmic rays are anisotropic has gained special attention in the last decade, since it could provide information about the galactic sources of the energetic particles and about the properties of the local interstellar medium and of the heliospheric magnetized plasma. Of particular interest is the evolution with energy of its angular structure, especially of the tail-in anisotropy which appears as a broad excess at sub-TeV energies from the direction of the heliotail, and seemingly degenerate into separate localized fractional excess regions above a few TeV. The directional coincidence of the tail-in excess at sub-TeV energies and of the most significant of the localized fractional excess regions at TeV energies, with the heliotail provides a compelling connection to this little known extended portion of the heliosphere.

Although we cannot exclude that other phenomena occur and might dominate the origin of the observation, such as the effect of energetic cosmic ray interaction with the turbulent ripples along the heliotail, in this paper another mechanism is discussed. Namely that a fraction of cosmic rays propagating through the heliotail are re-accelerated via first-order Fermi acceleration mechanism in weakly stochastic magnetic reconnection processes that originate in sectored magnetic field domains produced by the 11-year solar cycle. In general 3D numerical simulation show that such an acceleration mechanism can be efficient up to a few TeV, where a flatter than average spectrum could arise, depending on the competing effects of multiple acceleration processes and escape or loss effects, and back-reaction. On the other hand the properties of magnetized plasma in the heliotail are not yet fully understood, therefore details of cosmic ray propagation in this region are still uncertain. Sub-TeV cosmic rays may be accelerated over extended regions and may undergo more scattering, thus producing a broader arrival distribution. While multi-TeV cosmic rays undergo more efficient acceleration and their localized substructure in arrival direction are more related to the acceleration sites along the heliotail. Such acceleration mechanism is intrinsically anisotropic and as long as scattering is sub-dominant it would generate a net energy gain that could explain the seemingly harder spectrum observed within the localized excess regions by Milagro. %The underlying global anisotropy of TeV cosmic rays, moreover, provides probable enhancements in the effects of the first order Fermi mechanisms in the magnetic reconnection regions.

Acceleration processes in weakly stochastic magnetic reconnection regions as described by~\cite{GL03} have been used in~\cite{lazopher} to explain the origin of the anomalous cosmic rays. The Voyager aircraft measurements show that the anomalous cosmic rays persist also downstream the termination shock, indicating that the site of their acceleration is within the heliosheath closer to the heliopause in the upstream interstellar flow direction. The sectored magnetic field arising from the 26 day solar rotation and originated by the difference between rotation and magnetic axes, are pushed away by the solar wind and compressed upstream toward the heliopause, causing magnetic reconnection and energetic particle acceleration. A similar model for the origin of anomalous cosmic rays was proposed by~\cite{acr2} where the process of collisionless reconnection was discussed. In this paper we discussed a similar mechanism of the re-acceleration of energetic cosmic ray particles, where the scale of the sectored magnetic field is significantly larger. The higher magnetic energy involved provides the possibility to accelerate higher energy particles in an observable manner in terms of a slightly anomalous energy spectrum and arrival distribution.

%A similar process was proposed to explain the origin of anomalous cosmic rays as due to acceleration in stochastic reconnection region within the heliosheath, produced by magnetic polarity changes induced by the 27-day solar rotation~\citep{lazopher, acr2}.

\begin{acknowledgements}
The authors would like to thank the reviewers who helped to improve the paper. AL acknowledges the support of the NSF grant AST 0808118, NASA grant X5166204101 and of the NSF-sponsored Center for Magnetic Self-Organization. PD acknowledges the support from the U.S. National Science Foundation-Office of Polar Programs.
\end{acknowledgements}

\bibliography{papers}

\begin{thebibliography}{81}
\providecommand{\natexlab}[1]{#1}
\providecommand{\url}[1]{{\tt #1}}
\providecommand{\urlprefix}{URL }
\expandafter\ifx\csname urlstyle\endcsname\relax
  \providecommand{\doi}[1]{doi:\discretionary{}{}{}#1}\else
  \providecommand{\doi}{doi:\discretionary{}{}{}\begingroup
  \urlstyle{rm}\Url}\fi

\bibitem[{{Abbasi} et~al.(2010){Abbasi}, {Abdou}, {Abu-Zayyad}, {Adams},
  {Aguilar}, {Ahlers}, {Andeen}, {Auffenberg}, {Bai}, {Baker}, and
  et~al.}]{abbasi}
{Abbasi}, R., {Abdou}, Y., {Abu-Zayyad}, T., {Adams}, J., {Aguilar}, J.~A.,
  {Ahlers}, M., {Andeen}, K., {Auffenberg}, J., {Bai}, X., {Baker}, M., and
  et~al.: {Measurement of the Anisotropy of Cosmic-ray Arrival Directions with
  IceCube}, Astrophys. J. Lett., 718, L194--L198,
  \doi{10.1088/2041-8205/718/2/L194}, 2010.

\bibitem[{{Abbasi} et~al.(2011{\natexlab{a}}){Abbasi}, {Abdou}, {Abu-Zayyad},
  {Ackermann}, {Adams}, {Aguilar}, {Ahlers}, {Allen}, {Altmann}, and
  et~al.}]{abbasi11}
{Abbasi}, R., {Abdou}, Y., {Abu-Zayyad}, T., {Ackermann}, M., {Adams}, J.,
  {Aguilar}, J.~A., {Ahlers}, M., {Allen}, M.~M., {Altmann}, D., and et~al.:
  {Observation of an Anisotropy in the Galactic Cosmic Ray arrival direction at
  400 TeV with IceCube}, arXiv:1109.1017 [hep-ex], 2011{\natexlab{a}}.

\bibitem[{{Abbasi} et~al.(2011{\natexlab{b}}){Abbasi}, {Abdou}, {Abu-Zayyad},
  {Adams}, {Aguilar}, {Ahlers}, {Altmann}, {Andeen}, {Auffenberg}, {Bai}, and
  et~al.}]{abbasi11b}
{Abbasi}, R., {Abdou}, Y., {Abu-Zayyad}, T., {Adams}, J., {Aguilar}, J.~A.,
  {Ahlers}, M., {Altmann}, D., {Andeen}, K., {Auffenberg}, J., {Bai}, X., and
  et~al.: {Observation of Anisotropy in the Arrival Directions of Galactic
  Cosmic Rays at Multiple Angular Scales with IceCube}, Astrophys. J., 740,
  16--+, \doi{10.1088/0004-637X/740/1/16}, 2011{\natexlab{b}}.

\bibitem[{{Abdo} et~al.(2008){Abdo}, {Allen}, {Aune}, {Berley}, {Blaufuss},
  {Casanova}, {Chen}, {Dingus}, {Ellsworth}, {Fleysher}, {Fleysher},
  {Gonzalez}, {Goodman}, {Hoffman}, {H{\"u}ntemeyer}, {Kolterman}, {Lansdell},
  {Linnemann}, {McEnery}, {Mincer}, {Nemethy}, {Noyes}, {Pretz}, {Ryan},
  {Parkinson}, {Shoup}, {Sinnis}, {Smith}, {Sullivan}, {Vasileiou}, {Walker},
  {Williams}, and {Yodh}}]{abdo2}
{Abdo}, A.~A., {Allen}, B., {Aune}, T., {Berley}, D., {Blaufuss}, E.,
  {Casanova}, S., {Chen}, C., {Dingus}, B.~L., {Ellsworth}, R.~W., {Fleysher},
  L., {Fleysher}, R., {Gonzalez}, M.~M., {Goodman}, J.~A., {Hoffman}, C.~M.,
  {H{\"u}ntemeyer}, P.~H., {Kolterman}, B.~E., {Lansdell}, C.~P., {Linnemann},
  J.~T., {McEnery}, J.~E., {Mincer}, A.~I., {Nemethy}, P., {Noyes}, D.,
  {Pretz}, J., {Ryan}, J.~M., {Parkinson}, P.~M.~S., {Shoup}, A., {Sinnis}, G.,
  {Smith}, A.~J., {Sullivan}, G.~W., {Vasileiou}, V., {Walker}, G.~P.,
  {Williams}, D.~A., and {Yodh}, G.~B.: {Discovery of Localized Regions of
  Excess 10-TeV Cosmic Rays}, Physical Review Letters, 101, 221\,101--+,
  \doi{10.1103/PhysRevLett.101.221101}, 2008.

\bibitem[{{Abdo} et~al.(2009){Abdo}, {Allen}, {Aune}, {Berley}, {Casanova},
  {Chen}, {Dingus}, {Ellsworth}, {Fleysher}, {Fleysher}, {Gonzalez}, {Goodman},
  {Hoffman}, {Hopper}, {H{\"u}ntemeyer}, {Kolterman}, {Lansdell}, {Linnemann},
  {McEnery}, {Mincer}, {Nemethy}, {Noyes}, {Pretz}, {Ryan}, {Parkinson},
  {Shoup}, {Sinnis}, {Smith}, {Sullivan}, {Vasileiou}, {Walker}, {Williams},
  and {Yodh}}]{abdo}
{Abdo}, A.~A., {Allen}, B.~T., {Aune}, T., {Berley}, D., {Casanova}, S.,
  {Chen}, C., {Dingus}, B.~L., {Ellsworth}, R.~W., {Fleysher}, L., {Fleysher},
  R., {Gonzalez}, M.~M., {Goodman}, J.~A., {Hoffman}, C.~M., {Hopper}, B.,
  {H{\"u}ntemeyer}, P.~H., {Kolterman}, B.~E., {Lansdell}, C.~P., {Linnemann},
  J.~T., {McEnery}, J.~E., {Mincer}, A.~I., {Nemethy}, P., {Noyes}, D.,
  {Pretz}, J., {Ryan}, J.~M., {Parkinson}, P.~M.~S., {Shoup}, A., {Sinnis}, G.,
  {Smith}, A.~J., {Sullivan}, G.~W., {Vasileiou}, V., {Walker}, G.~P.,
  {Williams}, D.~A., and {Yodh}, G.~B.: {The Large-Scale Cosmic-Ray Anisotropy
  as Observed with Milagro}, Astrophys. J., 698, 2121--2130,
  \doi{10.1088/0004-637X/698/2/2121}, 2009.

\bibitem[{{Adriani} et~al.(2011){Adriani}, {Barbarino}, {Bazilevskaya},
  {Bellotti}, {Boezio}, {Bogomolov}, {Bonechi}, {Bongi}, {Bonvicini},
  {Borisov}, {Bottai}, {Bruno}, {Cafagna}, {Campana}, {Carbone}, {Carlson},
  {Casolino}, {Castellini}, {Consiglio}, {De Pascale}, {De Santis}, {De
  Simone}, {Di Felice}, {Galper}, {Gillard}, {Grishantseva}, {Jerse},
  {Karelin}, {Koldashov}, {Krutkov}, {Kvashnin}, {Leonov}, {Malakhov},
  {Malvezzi}, {Marcelli}, {Mayorov}, {Menn}, {Mikhailov}, {Mocchiutti},
  {Monaco}, {Mori}, {Nikonov}, {Osteria}, {Palma}, {Papini}, {Pearce},
  {Picozza}, {Pizzolotto}, {Ricci}, {Ricciarini}, {Rossetto}, {Sarkar},
  {Simon}, {Sparvoli}, {Spillantini}, {Stozhkov}, {Vacchi}, {Vannuccini},
  {Vasilyev}, {Voronov}, {Yurkin}, {Wu}, {Zampa}, {Zampa}, and
  {Zverev}}]{pamela}
{Adriani}, O., {Barbarino}, G.~C., {Bazilevskaya}, G.~A., {Bellotti}, R.,
  {Boezio}, M., {Bogomolov}, E.~A., {Bonechi}, L., {Bongi}, M., {Bonvicini},
  V., {Borisov}, S., {Bottai}, S., {Bruno}, A., {Cafagna}, F., {Campana}, D.,
  {Carbone}, R., {Carlson}, P., {Casolino}, M., {Castellini}, G., {Consiglio},
  L., {De Pascale}, M.~P., {De Santis}, C., {De Simone}, N., {Di Felice}, V.,
  {Galper}, A.~M., {Gillard}, W., {Grishantseva}, L., {Jerse}, G., {Karelin},
  A.~V., {Koldashov}, S.~V., {Krutkov}, S.~Y., {Kvashnin}, A.~N., {Leonov}, A.,
  {Malakhov}, V., {Malvezzi}, V., {Marcelli}, L., {Mayorov}, A.~G., {Menn}, W.,
  {Mikhailov}, V.~V., {Mocchiutti}, E., {Monaco}, A., {Mori}, N., {Nikonov},
  N., {Osteria}, G., {Palma}, F., {Papini}, P., {Pearce}, M., {Picozza}, P.,
  {Pizzolotto}, C., {Ricci}, M., {Ricciarini}, S.~B., {Rossetto}, L., {Sarkar},
  R., {Simon}, M., {Sparvoli}, R., {Spillantini}, P., {Stozhkov}, Y.~I.,
  {Vacchi}, A., {Vannuccini}, E., {Vasilyev}, G., {Voronov}, S.~A., {Yurkin},
  Y.~T., {Wu}, J., {Zampa}, G., {Zampa}, N., and {Zverev}, V.~G.: {PAMELA
  Measurements of Cosmic-Ray Proton and Helium Spectra}, Science, 332, 69--,
  \doi{10.1126/science.1199172}, 2011.

\bibitem[{Aglietta et~al.(2009)Aglietta, Alekseenko, Alessandro, Antonioli,
  Arneodo, Bergamasco, Bertaina, Bonino, Castellina, Chiavassa, Piazzoli,
  Sciascio, Fulgione, Galeotti, Ghia, Iacovacci, Mannocchi, Morello, Navarra,
  Saavedra, Stamerra, Trinchero, Valchierotti, Vallania, Vernetto, and
  Vigorito}]{aglietta}
Aglietta, M., Alekseenko, V.~V., Alessandro, B., Antonioli, P., Arneodo, F.,
  Bergamasco, L., Bertaina, M., Bonino, R., Castellina, A., Chiavassa, A.,
  Piazzoli, B.~D., Sciascio, G.~D., Fulgione, W., Galeotti, P., Ghia, P.~L.,
  Iacovacci, M., Mannocchi, G., Morello, C., Navarra, G., Saavedra, O.,
  Stamerra, A., Trinchero, G.~C., Valchierotti, S., Vallania, P., Vernetto, S.,
  and Vigorito, C.: Evolution of the Cosmic-Ray Anisotropy Above 1014 eV,
  Astrophys. J. Lett., 692, L130,
  \urlprefix\url{http://stacks.iop.org/1538-4357/692/i=2/a=L130}, 2009.

\bibitem[{{Ahn} et~al.(2010){Ahn}, {Allison}, {Bagliesi}, {Beatty},
  {Bigongiari}, {Childers}, {Conklin}, {Coutu}, {DuVernois}, {Ganel}, {Han},
  {Jeon}, {Kim}, {Lee}, {Lutz}, {Maestro}, {Malinin}, {Marrocchesi}, {Minnick},
  {Mognet}, {Nam}, {Nam}, {Nutter}, {Park}, {Park}, {Seo}, {Sina}, {Wu},
  {Yang}, {Yoon}, {Zei}, and {Zinn}}]{cream}
{Ahn}, H.~S., {Allison}, P., {Bagliesi}, M.~G., {Beatty}, J.~J., {Bigongiari},
  G., {Childers}, J.~T., {Conklin}, N.~B., {Coutu}, S., {DuVernois}, M.~A.,
  {Ganel}, O., {Han}, J.~H., {Jeon}, J.~A., {Kim}, K.~C., {Lee}, M.~H., {Lutz},
  L., {Maestro}, P., {Malinin}, A., {Marrocchesi}, P.~S., {Minnick}, S.,
  {Mognet}, S.~I., {Nam}, J., {Nam}, S., {Nutter}, S.~L., {Park}, I.~H.,
  {Park}, N.~H., {Seo}, E.~S., {Sina}, R., {Wu}, J., {Yang}, J., {Yoon}, Y.~S.,
  {Zei}, R., and {Zinn}, S.~Y.: {Discrepant Hardening Observed in Cosmic-ray
  Elemental Spectra}, Astrophys. J. Lett., 714, L89--L93,
  \doi{10.1088/2041-8205/714/1/L89}, 2010.

\bibitem[{{Amenomori} et~al.(2006){Amenomori}, {Ayabe}, {Bi}, {Chen}, {Cui},
  {Danzengluobu}, {Ding}, {Ding}, {Feng}, {Feng}, {Feng}, {Gao}, {Geng}, {Guo},
  {He}, {He}, {Hibino}, {Hotta}, {Hu}, {Hu}, {Huang}, {Huang}, {Jia}, {Kajino},
  {Kasahara}, {Katayose}, {Kato}, {Kawata}, {Labaciren}, {Le}, {Li}, {Li},
  {Lou}, {Lu}, {Lu}, {Meng}, {Mizutani}, {Mu}, {Munakata}, {Nagai}, {Nanjo},
  {Nishizawa}, {Ohnishi}, {Ohta}, {Onuma}, {Ouchi}, {Ozawa}, {Ren}, {Saito},
  {Saito}, {Sakata}, {Sako}, {Sasaki}, {Shibata}, {Shiomi}, {Shirai},
  {Sugimoto}, {Takita}, {Tan}, {Tateyama}, {Torii}, {Tsuchiya}, {Udo}, {Wang},
  {Wang}, {Wang}, {Wang}, {Wu}, {Xue}, {Yamamoto}, {Yan}, {Yang}, {Yasue},
  {Ye}, {Yu}, {Yuan}, {Yuda}, {Zhang}, {Zhang}, {Zhang}, {Zhang}, {Zhang},
  {Zhang}, {Zhaxisangzhu}, {Zhou}, and {The Tibet AS{$\gamma$}
  Collaboration}}]{amenomori}
{Amenomori}, M., {Ayabe}, S., {Bi}, X.~J., {Chen}, D., {Cui}, S.~W.,
  {Danzengluobu}, {Ding}, L.~K., {Ding}, X.~H., {Feng}, C.~F., {Feng}, Z.,
  {Feng}, Z.~Y., {Gao}, X.~Y., {Geng}, Q.~X., {Guo}, H.~W., {He}, H.~H., {He},
  M., {Hibino}, K., {Hotta}, N., {Hu}, H., {Hu}, H.~B., {Huang}, J., {Huang},
  Q., {Jia}, H.~Y., {Kajino}, F., {Kasahara}, K., {Katayose}, Y., {Kato}, C.,
  {Kawata}, K., {Labaciren}, {Le}, G.~M., {Li}, A.~F., {Li}, J.~Y., {Lou},
  Y.-Q., {Lu}, H., {Lu}, S.~L., {Meng}, X.~R., {Mizutani}, K., {Mu}, J.,
  {Munakata}, K., {Nagai}, A., {Nanjo}, H., {Nishizawa}, M., {Ohnishi}, M.,
  {Ohta}, I., {Onuma}, H., {Ouchi}, T., {Ozawa}, S., {Ren}, J.~R., {Saito}, T.,
  {Saito}, T.~Y., {Sakata}, M., {Sako}, T.~K., {Sasaki}, T., {Shibata}, M.,
  {Shiomi}, A., {Shirai}, T., {Sugimoto}, H., {Takita}, M., {Tan}, Y.~H.,
  {Tateyama}, N., {Torii}, S., {Tsuchiya}, H., {Udo}, S., {Wang}, B., {Wang},
  H., {Wang}, X., {Wang}, Y.~G., {Wu}, H.~R., {Xue}, L., {Yamamoto}, Y., {Yan},
  C.~T., {Yang}, X.~C., {Yasue}, S., {Ye}, Z.~H., {Yu}, G.~C., {Yuan}, A.~F.,
  {Yuda}, T., {Zhang}, H.~M., {Zhang}, J.~L., {Zhang}, N.~J., {Zhang}, X.~Y.,
  {Zhang}, Y., {Zhang}, Y., {Zhaxisangzhu}, {Zhou}, X.~X., and {The Tibet
  AS{$\gamma$} Collaboration}: {Anisotropy and Corotation of Galactic Cosmic
  Rays}, Science, 314, 439--443, \doi{10.1126/science.1131702}, 2006.

\bibitem[{{Amenomori} et~al.(2007){Amenomori}, {Ayabe}, {Bi}, {Chen}, {Cui},
  {Danzengluobu}, {Ding}, {Ding}, {Feng}, {Feng}, {Feng}, {Gao}, {Geng}, {Guo},
  {He}, {He}, {Hibino}, {Hotta}, {Hu}, {Hu}, {Huang}, {Huang}, {Jia}, {Kajino},
  {Kasahara}, {Katayose}, {Kato}, {Kawata}, {Labaciren}, {Le}, {Li}, {Li},
  {Lou}, {Lu}, {Lu}, {Meng}, {Mizutani}, {Mu}, {Munakata}, {Nagai}, {Nanjo},
  {Nishizawa}, {Ohnishi}, {Ohta}, {Onuma}, {Ouchi}, {Ozawa}, {Ren}, {Saito},
  {Saito}, {Sakata}, {Sako}, {Sasaki}, {Shibata}, {Shiomi}, {Shirai},
  {Sugimoto}, {Takita}, {Tan}, {Tateyama}, {Torii}, {Tsuchiya}, {Udo}, {Wang},
  {Wang}, {Wang}, {Wang}, {Wu}, {Xue}, {Yamamoto}, {Yan}, {Yang}, {Yasue},
  {Ye}, {Yu}, {Yuan}, {Yuda}, {Zhang}, {Zhang}, {Zhang}, {Zhang}, {Zhang},
  {Zhang}, {Zhaxisangzhu}, {Zhou}, and {The Tibet AS{$\gamma$}
  Collaboration}}]{amenomori2}
{Amenomori}, M., {Ayabe}, S., {Bi}, X.~J., {Chen}, D., {Cui}, S.~W.,
  {Danzengluobu}, {Ding}, L.~K., {Ding}, X.~H., {Feng}, C.~F., {Feng}, Z.,
  {Feng}, Z.~Y., {Gao}, X.~Y., {Geng}, Q.~X., {Guo}, H.~W., {He}, H.~H., {He},
  M., {Hibino}, K., {Hotta}, N., {Hu}, H., {Hu}, H.~B., {Huang}, J., {Huang},
  Q., {Jia}, H.~Y., {Kajino}, F., {Kasahara}, K., {Katayose}, Y., {Kato}, C.,
  {Kawata}, K., {Labaciren}, {Le}, G.~M., {Li}, A.~F., {Li}, J.~Y., {Lou},
  Y.-Q., {Lu}, H., {Lu}, S.~L., {Meng}, X.~R., {Mizutani}, K., {Mu}, J.,
  {Munakata}, K., {Nagai}, A., {Nanjo}, H., {Nishizawa}, M., {Ohnishi}, M.,
  {Ohta}, I., {Onuma}, H., {Ouchi}, T., {Ozawa}, S., {Ren}, J.~R., {Saito}, T.,
  {Saito}, T.~Y., {Sakata}, M., {Sako}, T.~K., {Sasaki}, T., {Shibata}, M.,
  {Shiomi}, A., {Shirai}, T., {Sugimoto}, H., {Takita}, M., {Tan}, Y.~H.,
  {Tateyama}, N., {Torii}, S., {Tsuchiya}, H., {Udo}, S., {Wang}, B., {Wang},
  H., {Wang}, X., {Wang}, Y.~G., {Wu}, H.~R., {Xue}, L., {Yamamoto}, Y., {Yan},
  C.~T., {Yang}, X.~C., {Yasue}, S., {Ye}, Z.~H., {Yu}, G.~C., {Yuan}, A.~F.,
  {Yuda}, T., {Zhang}, H.~M., {Zhang}, J.~L., {Zhang}, N.~J., {Zhang}, X.~Y.,
  {Zhang}, Y., {Zhang}, Y., {Zhaxisangzhu}, {Zhou}, X.~X., and {The Tibet
  AS{$\gamma$} Collaboration}: {Implication of the sidereal anisotropy of \~{}5
  TeV cosmic ray intensity observed with the Tibet III air shower array}, in:
  International Cosmic Ray Conference, vol.~1 of {\em International Cosmic Ray
  Conference, M\'erida, Mexico\/}, 2007.

\bibitem[{{Amenomori} et~al.(2011{\natexlab{a}}){Amenomori}, {Ayabe}, {Bi},
  {Chen}, {Cui}, {Danzengluobu}, {Ding}, {Ding}, {Feng}, {Feng}, {Feng}, {Gao},
  {Geng}, {Guo}, {He}, {He}, {Hibino}, {Hotta}, {Hu}, {Hu}, {Huang}, {Huang},
  {Jia}, {Kajino}, {Kasahara}, {Katayose}, {Kato}, {Kawata}, {Labaciren}, {Le},
  {Li}, {Li}, {Lou}, {Lu}, {Lu}, {Meng}, {Mizutani}, {Mu}, {Munakata}, {Nagai},
  {Nanjo}, {Nishizawa}, {Ohnishi}, {Ohta}, {Onuma}, {Ouchi}, {Ozawa}, {Ren},
  {Saito}, {Saito}, {Sakata}, {Sako}, {Sasaki}, {Shibata}, {Shiomi}, {Shirai},
  {Sugimoto}, {Takita}, {Tan}, {Tateyama}, {Torii}, {Tsuchiya}, {Udo}, {Wang},
  {Wang}, {Wang}, {Wang}, {Wu}, {Xue}, {Yamamoto}, {Yan}, {Yang}, {Yasue},
  {Ye}, {Yu}, {Yuan}, {Yuda}, {Zhang}, {Zhang}, {Zhang}, {Zhang}, {Zhang},
  {Zhang}, {Zhaxisangzhu}, {Zhou}, and {The Tibet AS{$\gamma$}
  Collaboration}}]{amenomori11}
{Amenomori}, M., {Ayabe}, S., {Bi}, X.~J., {Chen}, D., {Cui}, S.~W.,
  {Danzengluobu}, {Ding}, L.~K., {Ding}, X.~H., {Feng}, C.~F., {Feng}, Z.,
  {Feng}, Z.~Y., {Gao}, X.~Y., {Geng}, Q.~X., {Guo}, H.~W., {He}, H.~H., {He},
  M., {Hibino}, K., {Hotta}, N., {Hu}, H., {Hu}, H.~B., {Huang}, J., {Huang},
  Q., {Jia}, H.~Y., {Kajino}, F., {Kasahara}, K., {Katayose}, Y., {Kato}, C.,
  {Kawata}, K., {Labaciren}, {Le}, G.~M., {Li}, A.~F., {Li}, J.~Y., {Lou},
  Y.-Q., {Lu}, H., {Lu}, S.~L., {Meng}, X.~R., {Mizutani}, K., {Mu}, J.,
  {Munakata}, K., {Nagai}, A., {Nanjo}, H., {Nishizawa}, M., {Ohnishi}, M.,
  {Ohta}, I., {Onuma}, H., {Ouchi}, T., {Ozawa}, S., {Ren}, J.~R., {Saito}, T.,
  {Saito}, T.~Y., {Sakata}, M., {Sako}, T.~K., {Sasaki}, T., {Shibata}, M.,
  {Shiomi}, A., {Shirai}, T., {Sugimoto}, H., {Takita}, M., {Tan}, Y.~H.,
  {Tateyama}, N., {Torii}, S., {Tsuchiya}, H., {Udo}, S., {Wang}, B., {Wang},
  H., {Wang}, X., {Wang}, Y.~G., {Wu}, H.~R., {Xue}, L., {Yamamoto}, Y., {Yan},
  C.~T., {Yang}, X.~C., {Yasue}, S., {Ye}, Z.~H., {Yu}, G.~C., {Yuan}, A.~F.,
  {Yuda}, T., {Zhang}, H.~M., {Zhang}, J.~L., {Zhang}, N.~J., {Zhang}, X.~Y.,
  {Zhang}, Y., {Zhang}, Y., {Zhaxisangzhu}, {Zhou}, X.~X., and {The Tibet
  AS{$\gamma$} Collaboration}: {Time Dependence of Loss-Cone Amplitude measured
  with the Tibet Air-Shower Array}, in: International Cosmic Ray Conference,
  vol.~1 of {\em International Cosmic Ray Conference, Beijing China\/}, p.
  0379, 2011{\natexlab{a}}.

\bibitem[{{Amenomori} et~al.(2011{\natexlab{b}}){Amenomori}, {Ayabe}, {Bi},
  {Chen}, {Cui}, {Danzengluobu}, {Ding}, {Ding}, {Feng}, {Feng}, {Feng}, {Gao},
  {Geng}, {Guo}, {He}, {He}, {Hibino}, {Hotta}, {Hu}, {Hu}, {Huang}, {Huang},
  {Jia}, {Kajino}, {Kasahara}, {Katayose}, {Kato}, {Kawata}, {Labaciren}, {Le},
  {Li}, {Li}, {Lou}, {Lu}, {Lu}, {Meng}, {Mizutani}, {Mu}, {Munakata}, {Nagai},
  {Nanjo}, {Nishizawa}, {Ohnishi}, {Ohta}, {Onuma}, {Ouchi}, {Ozawa}, {Ren},
  {Saito}, {Saito}, {Sakata}, {Sako}, {Sasaki}, {Shibata}, {Shiomi}, {Shirai},
  {Sugimoto}, {Takita}, {Tan}, {Tateyama}, {Torii}, {Tsuchiya}, {Udo}, {Wang},
  {Wang}, {Wang}, {Wang}, {Wu}, {Xue}, {Yamamoto}, {Yan}, {Yang}, {Yasue},
  {Ye}, {Yu}, {Yuan}, {Yuda}, {Zhang}, {Zhang}, {Zhang}, {Zhang}, {Zhang},
  {Zhang}, {Zhaxisangzhu}, {Zhou}, and {The Tibet AS{$\gamma$}
  Collaboration}}]{amenomori11b}
{Amenomori}, M., {Ayabe}, S., {Bi}, X.~J., {Chen}, D., {Cui}, S.~W.,
  {Danzengluobu}, {Ding}, L.~K., {Ding}, X.~H., {Feng}, C.~F., {Feng}, Z.,
  {Feng}, Z.~Y., {Gao}, X.~Y., {Geng}, Q.~X., {Guo}, H.~W., {He}, H.~H., {He},
  M., {Hibino}, K., {Hotta}, N., {Hu}, H., {Hu}, H.~B., {Huang}, J., {Huang},
  Q., {Jia}, H.~Y., {Kajino}, F., {Kasahara}, K., {Katayose}, Y., {Kato}, C.,
  {Kawata}, K., {Labaciren}, {Le}, G.~M., {Li}, A.~F., {Li}, J.~Y., {Lou},
  Y.-Q., {Lu}, H., {Lu}, S.~L., {Meng}, X.~R., {Mizutani}, K., {Mu}, J.,
  {Munakata}, K., {Nagai}, A., {Nanjo}, H., {Nishizawa}, M., {Ohnishi}, M.,
  {Ohta}, I., {Onuma}, H., {Ouchi}, T., {Ozawa}, S., {Ren}, J.~R., {Saito}, T.,
  {Saito}, T.~Y., {Sakata}, M., {Sako}, T.~K., {Sasaki}, T., {Shibata}, M.,
  {Shiomi}, A., {Shirai}, T., {Sugimoto}, H., {Takita}, M., {Tan}, Y.~H.,
  {Tateyama}, N., {Torii}, S., {Tsuchiya}, H., {Udo}, S., {Wang}, B., {Wang},
  H., {Wang}, X., {Wang}, Y.~G., {Wu}, H.~R., {Xue}, L., {Yamamoto}, Y., {Yan},
  C.~T., {Yang}, X.~C., {Yasue}, S., {Ye}, Z.~H., {Yu}, G.~C., {Yuan}, A.~F.,
  {Yuda}, T., {Zhang}, H.~M., {Zhang}, J.~L., {Zhang}, N.~J., {Zhang}, X.~Y.,
  {Zhang}, Y., {Zhang}, Y., {Zhaxisangzhu}, {Zhou}, X.~X., and {The Tibet
  AS{$\gamma$} Collaboration}: {Modeling of the galactic cosmic-ray anisotropy
  at TeV energies}, in: International Cosmic Ray Conference, vol.~1 of {\em
  International Cosmic Ray Conference, Beijing, China\/}, p. 0361,
  2011{\natexlab{b}}.

\bibitem[{{Battaner} et~al.(2009){Battaner}, {Castellano}, and
  {Masip}}]{battaner}
{Battaner}, E., {Castellano}, J., and {Masip}, M.: {Galactic Magnetic Fields
  and the Large-Scale Anisotropy at Milagro}, Astrophys. J. Lett., 703,
  L90--L93, \doi{10.1088/0004-637X/703/1/L90}, 2009.

\bibitem[{{Bell}(1978)}]{bell}
{Bell}, A.~R.: {The acceleration of cosmic rays in shock fronts. II}, MNRAS,
  182, 443--455, 1978.

\bibitem[{{Beresnyak} et~al.(2011){Beresnyak}, {Yan}, and {Lazarian}}]{beres11}
{Beresnyak}, A., {Yan}, H., and {Lazarian}, A.: {Numerical Study of Cosmic Ray
  Diffusion in Magnetohydrodynamic Turbulence}, Astrophys. J., 728, 60--+,
  \doi{10.1088/0004-637X/728/1/60}, 2011.

\bibitem[{{Birn} et~al.(2004){Birn}, {Thomsen}, and {Hesse}}]{birn}
{Birn}, J., {Thomsen}, M.~F., and {Hesse}, M.: {Electron acceleration in the
  dynamic magnetotail: Test particle orbits in three-dimensional
  magnetohydrodynamic simulation fields}, Physics of Plasmas, 11, 1825--1833,
  \doi{10.1063/1.1704641}, 2004.

\bibitem[{{Biskamp}(1996)}]{biskamp96}
{Biskamp}, D.: {Magnetic Reconnection in Plasmas}, Astrophys. and Space
  Science, 242, 165--207, \doi{10.1007/BF00645113}, 1996.

\bibitem[{{Blasi} and {Amato}(2012)}]{blasi}
{Blasi}, P. and {Amato}, E.: {Diffusive propagation of cosmic rays from
  supernova remnants in the Galaxy. II: anisotropy}, JCAP, 1, 11,
  \doi{10.1088/1475-7516/2012/01/011}, 2012.

\bibitem[{{Ciaravella} and {Raymond}(2008)}]{ciaravella}
{Ciaravella}, A. and {Raymond}, J.~C.: {The Current Sheet Associated with the
  2003 November 4 Coronal Mass Ejection: Density, Temperature, Thickness, and
  Line Width}, Astrophys. J., 686, 1372--1382, \doi{10.1086/590655}, 2008.

\bibitem[{{Compton} and {Getting}(1935)}]{compton}
{Compton}, A.~H. and {Getting}, I.~A.: {An Apparent Effect of Galactic Rotation
  on the Intensity of Cosmic Rays}, Physical Review, 47, 817--821,
  \doi{10.1103/PhysRev.47.817}, 1935.

\bibitem[{{de Gouveia Dal Pino} and {Lazarian}(2003)}]{GL03}
{de Gouveia Dal Pino}, E.~M. and {Lazarian}, A.: {The role of Violent Magnetic
  Reconnection on the Production of the Large Scale Superluminal Ejections of
  the Microquasar GRS 1915+105}, arXiv:astro-ph/0307054, 2003.

\bibitem[{{de Gouveia Dal Pino} and {Lazarian}(2005)}]{gouveia}
{de Gouveia Dal Pino}, E.~M. and {Lazarian}, A.: {Production of the large scale
  superluminal ejections of the microquasar GRS 1915+105 by violent magnetic
  reconnection}, A\&A, 441, 845--853, \doi{10.1051/0004-6361:20042590}, 2005.

\bibitem[{{Desiati} and {Lazarian}(2011)}]{locscat}
{Desiati}, P. and {Lazarian}, A.: {Anisotropy of TeV Cosmic Rays and the Outer
  Heliospheric Boundaries}, ArXiv e-prints, 2011.

\bibitem[{{Di Sciascio} and {for the ARGO-YBJ Collaboration}(2012)}]{argo12}
{Di Sciascio}, G. and {for the ARGO-YBJ Collaboration}: {Measurement of Cosmic
  Ray spectrum and Anisotropy with ARGO-YBJ}, ArXiv e-prints, 2012.

\bibitem[{{Drake} et~al.(2006){Drake}, {Swisdak}, {Che}, and {Shay}}]{drake}
{Drake}, J.~F., {Swisdak}, M., {Che}, H., and {Shay}, M.~A.: {Electron
  acceleration from contracting magnetic islands during reconnection}, Nature,
  443, 553--556, \doi{10.1038/nature05116}, 2006.

\bibitem[{{Drake} et~al.(2010){Drake}, {Opher}, {Swisdak}, and
  {Chamoun}}]{acr2}
{Drake}, J.~F., {Opher}, M., {Swisdak}, M., and {Chamoun}, J.~N.: {A Magnetic
  Reconnection Mechanism for the Generation of Anomalous Cosmic Rays},
  Astrophys. J., 709, 963--974, \doi{10.1088/0004-637X/709/2/963}, 2010.

\bibitem[{{Drury} and {Aharonian}(2008)}]{drury}
{Drury}, L.~O.~. and {Aharonian}, F.~A.: {The puzzling MILAGRO hot spots},
  Astroparticle Physics, 29, 420--423,
  \doi{10.1016/j.astropartphys.2008.04.007}, 2008.

\bibitem[{{Erlykin} and {Wolfendale}(2006)}]{erlykin}
{Erlykin}, A.~D. and {Wolfendale}, A.~W.: {The anisotropy of galactic cosmic
  rays as a product of stochastic supernova explosions}, Astroparticle Physics,
  25, 183--194, \doi{10.1016/j.astropartphys.2006.01.003}, 2006.

\bibitem[{{Frisch}(2011)}]{frisch2011b}
{Frisch}, P.~C.: {How Local is the Local Interstellar Magnetic Field?}, ArXiv
  e-prints, 2011.

\bibitem[{{Frisch} et~al.(2011){Frisch}, {Redfield}, and
  {Slavin}}]{frisch2011a}
{Frisch}, P.~C., {Redfield}, S., and {Slavin}, J.~D.: {The Interstellar Medium
  Surrounding the Sun}, Annu. Rev. Astro. Astrophys., 49, 237--279,
  \doi{10.1146/annurev-astro-081710-102613}, 2011.

\bibitem[{{Giacinti} and {Sigl}(2011)}]{giacinti}
{Giacinti}, G. and {Sigl}, G.: {Local Magnetic Turbulence and TeV-PeV Cosmic
  Ray Anisotropies}, ArXiv e-prints, 2011.

\bibitem[{{Gieseler} and {Jones}(2000)}]{gieseler}
{Gieseler}, U.~D.~J. and {Jones}, T.~W.: {First order Fermi acceleration at
  multiple oblique shocks}, A\&A, 357, 1133--1136, 2000.

\bibitem[{{Gleeson} and {Axford}(1968)}]{compton2}
{Gleeson}, L.~J. and {Axford}, W.~I.: {The Compton-Getting Effect}, Astrophys.
  and Space Science, 2, 431--437, \doi{10.1007/BF02175919}, 1968.

\bibitem[{{Guillian} et~al.(2007){Guillian}, {Hosaka}, {Ishihara}, {Kameda},
  {Koshio}, {Minamino}, {Mitsuda}, {Miura}, {Moriyama}, {Nakahata}, {Namba},
  {Obayashi}, {Ogawa}, {Shiozawa}, {Suzuki}, {Takeda}, {Takeuchi}, {Yamada},
  {Higuchi}, {Ishitsuka}, {Kajita}, {Kaneyuki}, {Mitsuka}, {Nakayama},
  {Nishino}, {Okada}, {Okumura}, {Saji}, {Takenaga}, {Desai}, {Kearns},
  {Stone}, {Sulak}, {Wang}, {Goldhaber}, {Casper}, {Gajewski}, {Griskevich},
  {Kropp}, {Liu}, {Mine}, {Smy}, {Sobel}, {Vagins}, {Ganezer}, {Hill}, {Keig},
  {Scholberg}, {Walter}, {Ellsworth}, {Tasaka}, {Kibayashi}, {Learned},
  {Matsuno}, {Messier}, {Hayato}, {Ichikawa}, {Ishida}, {Ishii}, {Iwashita},
  {Kobayashi}, {Nakadaira}, {Nakamura}, {Nitta}, {Oyama}, {Totsuka}, {Suzuki},
  {Hasegawa}, {Kato}, {Maesaka}, {Nakaya}, {Nishikawa}, {Sato}, {Yamamoto},
  {Yokoyama}, {Haines}, {Dazeley}, {Hatakeyama}, {Svoboda}, {Blaufuss},
  {Goodman}, {Sullivan}, {Turcan}, {Habig}, {Fukuda}, {Itow}, {Sakuda},
  {Yoshida}, {Kim}, {Yoo}, {Okazawa}, {Ishizuka}, {Jung}, {Kato}, {Kobayashi},
  {Malek}, {Mauger}, {McGrew}, {Sharkey}, {Yanagisawa}, {Gando}, {Hasegawa},
  {Inoue}, {Shirai}, {Suzuki}, {Nishijima}, {Ishino}, {Watanabe}, {Koshiba},
  {Kielczewska}, {Berns}, {Gran}, {Shiraishi}, {Stachyra}, {Washburn},
  {Wilkes}, and {Munakata}}]{guillian}
{Guillian}, G., {Hosaka}, J., {Ishihara}, K., {Kameda}, J., {Koshio}, Y.,
  {Minamino}, A., {Mitsuda}, C., {Miura}, M., {Moriyama}, S., {Nakahata}, M.,
  {Namba}, T., {Obayashi}, Y., {Ogawa}, H., {Shiozawa}, M., {Suzuki}, Y.,
  {Takeda}, A., {Takeuchi}, Y., {Yamada}, S., {Higuchi}, I., {Ishitsuka}, M.,
  {Kajita}, T., {Kaneyuki}, K., {Mitsuka}, G., {Nakayama}, S., {Nishino}, H.,
  {Okada}, A., {Okumura}, K., {Saji}, C., {Takenaga}, Y., {Desai}, S.,
  {Kearns}, E., {Stone}, J.~L., {Sulak}, L.~R., {Wang}, W., {Goldhaber}, M.,
  {Casper}, D., {Gajewski}, W., {Griskevich}, J., {Kropp}, W.~R., {Liu}, D.~W.,
  {Mine}, S., {Smy}, M.~B., {Sobel}, H.~W., {Vagins}, M.~R., {Ganezer}, K.~S.,
  {Hill}, J., {Keig}, W.~E., {Scholberg}, K., {Walter}, C.~W., {Ellsworth},
  R.~W., {Tasaka}, S., {Kibayashi}, A., {Learned}, J.~G., {Matsuno}, S.,
  {Messier}, M.~D., {Hayato}, Y., {Ichikawa}, A.~K., {Ishida}, T., {Ishii}, T.,
  {Iwashita}, T., {Kobayashi}, T., {Nakadaira}, T., {Nakamura}, K., {Nitta},
  K., {Oyama}, Y., {Totsuka}, Y., {Suzuki}, A.~T., {Hasegawa}, M., {Kato}, I.,
  {Maesaka}, H., {Nakaya}, T., {Nishikawa}, K., {Sato}, H., {Yamamoto}, S.,
  {Yokoyama}, M., {Haines}, T.~J., {Dazeley}, S., {Hatakeyama}, S., {Svoboda},
  R., {Blaufuss}, E., {Goodman}, J.~A., {Sullivan}, G.~W., {Turcan}, D.,
  {Habig}, A., {Fukuda}, Y., {Itow}, Y., {Sakuda}, M., {Yoshida}, M., {Kim},
  S.~B., {Yoo}, J., {Okazawa}, H., {Ishizuka}, T., {Jung}, C.~K., {Kato}, T.,
  {Kobayashi}, K., {Malek}, M., {Mauger}, C., {McGrew}, C., {Sharkey}, E.,
  {Yanagisawa}, C., {Gando}, Y., {Hasegawa}, T., {Inoue}, K., {Shirai}, J.,
  {Suzuki}, A., {Nishijima}, K., {Ishino}, H., {Watanabe}, Y., {Koshiba}, M.,
  {Kielczewska}, D., {Berns}, H.~G., {Gran}, R., {Shiraishi}, K.~K.,
  {Stachyra}, A.~L., {Washburn}, K., {Wilkes}, R.~J., and {Munakata}, K.:
  {Observation of the anisotropy of 10TeV primary cosmic ray nuclei flux with
  the Super-Kamiokande-I detector}, Phys. Rev. D, 75, 062\,003--+,
  \doi{10.1103/PhysRevD.75.062003}, 2007.

\bibitem[{{Iuppa}(2011)}]{iuppa11}
{Iuppa}, R.: {Few-degree anisotropies in the cosmic-ray flux observed by the
  ARGO-YBJ experiment}, in: International Cosmic Ray Conference, vol.~1 of {\em
  International Cosmic Ray Conference, Beijing China\/}, p. 0507, 2011.

\bibitem[{{Izmodenov} and {Alexashov}(2003)}]{izmodenov03}
{Izmodenov}, V.~V. and {Alexashov}, D.~B.: {A Model for the Tail Region of the
  Heliospheric Interface}, Astronomy Letters, 29, 58--63,
  \doi{10.1134/1.1537379}, 2003.

\bibitem[{{Izmodenov} and {Kallenbach}(2006)}]{izmodenov}
{Izmodenov}, V.~V. and {Kallenbach}, R., eds.: {The Physics of the Heliospheric
  Boundaries}, 2006.

\bibitem[{{Kliem}(1994)}]{kliem}
{Kliem}, B.: {Particle orbits, trapping, and acceleration in a filamentary
  current sheet model}, Astrophys. J. Suppl., 90, 719--728,
  \doi{10.1086/191896}, 1994.

\bibitem[{{Kowal} et~al.(2009){Kowal}, {Lazarian}, {Vishniac}, and
  {Otmianowska-Mazur}}]{kowal}
{Kowal}, G., {Lazarian}, A., {Vishniac}, E.~T., and {Otmianowska-Mazur}, K.:
  {Numerical Tests of Fast Reconnection in Weakly Stochastic Magnetic Fields},
  Astrophys. J., 700, 63--85, \doi{10.1088/0004-637X/700/1/63}, 2009.

\bibitem[{{Kowal} et~al.(2011){Kowal}, {de Gouveia Dal Pino}, and
  {Lazarian}}]{kowal11}
{Kowal}, G., {de Gouveia Dal Pino}, E.~M., and {Lazarian}, A.:
  {Magnetohydrodynamic Simulations of Reconnection and Particle Acceleration:
  Three-dimensional Effects}, Astrophys. J., 735, 102--+,
  \doi{10.1088/0004-637X/735/2/102}, 2011.

\bibitem[{{Kowal} et~al.(2012){Kowal}, {de Gouveia Dal Pino}, and
  {Lazarian}}]{kowal12}
{Kowal}, G., {de Gouveia Dal Pino}, E.~M., and {Lazarian}, A.: {Particle
  Acceleration in Turbulence and Weakly Stochastic Reconnection}, accepted for
  publication in PRL, 2012.

\bibitem[{{Lazarian}(2005)}]{laz05}
{Lazarian}, A.: {Magnetic Fields in the Universe: From Laboratory and Stars to
  Primordial Structures}, in: Magnetic Fields in the Universe: From Laboratory
  and Stars to Primordial Structures., edited by {de Gouveia Dal Pino}, E.~M.,
  {Lugones}, G., and {Lazarian}, A., vol. 784 of {\em American Institute of
  Physics Conference Series\/}, 2005.

\bibitem[{{Lazarian}(2006)}]{laz06}
{Lazarian}, A.: {Enhancement and Suppression of Heat Transfer by MHD
  Turbulence}, Astrophys. J. Lett., 645, L25--L28, \doi{10.1086/505796}, 2006.

\bibitem[{{Lazarian}(2007)}]{laz06err}
{Lazarian}, A.: {Erratum: ``Enhancement and Suppression of Heat Transfer by MHD
  Turbulence'' (<A href=''/abs/2006ApJ...645L..25L''>ApJ, 645, L25
  [2006]</A>)}, Astrophys. J. Lett., 660, L173--L173, \doi{10.1086/518163},
  2007.

\bibitem[{{Lazarian} and {Desiati}(2010)}]{reconnection}
{Lazarian}, A. and {Desiati}, P.: {Magnetic Reconnection as the Cause of Cosmic
  Ray Excess from the Heliospheric Tail}, Astrophys. J., 722, 188--196,
  \doi{10.1088/0004-637X/722/1/188}, 2010.

\bibitem[{{Lazarian} and {Opher}(2009)}]{lazopher}
{Lazarian}, A. and {Opher}, M.: {A Model of Acceleration of Anomalous Cosmic
  Rays by Reconnection in the Heliosheath}, Astrophys. J., 703, 8--21,
  \doi{10.1088/0004-637X/703/1/8}, 2009.

\bibitem[{{Lazarian} and {Vishniac}(1999)}]{lv99}
{Lazarian}, A. and {Vishniac}, E.~T.: {Reconnection in a Weakly Stochastic
  Field}, Astrophys. J., 517, 700--718, \doi{10.1086/307233}, 1999.

\bibitem[{{Lazarian} et~al.(2004){Lazarian}, {Vishniac}, and {Cho}}]{laz04}
{Lazarian}, A., {Vishniac}, E.~T., and {Cho}, J.: {Magnetic Field Structure and
  Stochastic Reconnection in a Partially Ionized Gas}, Astrophys. J., 603,
  180--197, \doi{10.1086/381383}, 2004.

\bibitem[{{Lazarian} et~al.(2011){Lazarian}, {Kowal}, {Vishniac}, and {de
  Gouveia Dal Pino}}]{laz11}
{Lazarian}, A., {Kowal}, G., {Vishniac}, E., and {de Gouveia Dal Pino}, E.:
  {Fast magnetic reconnection and energetic particle acceleration}, Plan. Sp.
  Science, 59, 537--546, \doi{10.1016/j.pss.2010.07.020}, 2011.

\bibitem[{{Liewer} et~al.(1996){Liewer}, {Karmesin}, and
  {Brackbill}}]{liewer96}
{Liewer}, P.~C., {Karmesin}, S.~R., and {Brackbill}, J.~U.: {Hydrodynamic
  instability of the heliopause driven by plasma-neutral charge-exchange
  interactions}, J. of Geophys. Res., 101, 17\,119--17\,128,
  \doi{10.1029/96JA00606}, 1996.

\bibitem[{{Longair}(1992)}]{longair}
{Longair}, M.~S.: {High energy astrophysics. Vol.2}, 1992.

\bibitem[{{Malkov} et~al.(2010){Malkov}, {Diamond}, {O'C.~Drury}, and
  {Sagdeev}}]{malkov}
{Malkov}, M.~A., {Diamond}, P.~H., {O'C.~Drury}, L., and {Sagdeev}, R.~Z.:
  {Probing Nearby Cosmic-ray Accelerators and Interstellar Medium Turbulence
  with MILAGRO Hot Spots}, Astrophys. J., 721, 750--761,
  \doi{10.1088/0004-637X/721/1/750}, 2010.

\bibitem[{{Matthaeus} et~al.(1984){Matthaeus}, {Ambrosiano}, and
  {Goldstein}}]{matthaeus}
{Matthaeus}, W.~H., {Ambrosiano}, J.~J., and {Goldstein}, M.~L.:
  {Particle-acceleration by turbulent magnetohydrodynamic reconnection},
  Physical Review Letters, 53, 1449--1452, \doi{10.1103/PhysRevLett.53.1449},
  1984.

\bibitem[{{Melrose} and {Pope}(1993)}]{melrose}
{Melrose}, D.~B. and {Pope}, M.~H.: {Diffusive Shock Acceleration by Multiple
  Shocks}, Proceedings of the Astronomical Society of Australia, 10, 222--+,
  1993.

\bibitem[{Munakata et~al.(2010)Munakata, Mizoguchi, Kato, Yasue, Mori, Takita,
  and K\'ota}]{munakata11}
Munakata, K., Mizoguchi, Y., Kato, C., Yasue, S., Mori, S., Takita, M., and
  K\'ota, J.: Solar Cycle Dependence of the Diurnal Anisotropy of 0.6 TeV
  Cosmic-ray Intensity Observed with the Matsushiro Underground Muon Detector,
  Astrophys. J., 712, 1100,
  \urlprefix\url{http://stacks.iop.org/0004-637X/712/i=2/a=1100}, 2010.

\bibitem[{{Nagashima} et~al.(1998){Nagashima}, {Fujimoto}, and
  {Jacklyn}}]{nagashima}
{Nagashima}, K., {Fujimoto}, K., and {Jacklyn}, R.~M.: {Galactic and
  heliotail-in anisotropies of cosmic rays as the origin of sidereal daily
  variation in the energy region <10$^{4}$GeV}, J. of Geophys. Res., 1031,
  17\,429--17\,440, \doi{10.1029/98JA01105}, 1998.

\bibitem[{{Nerney} et~al.(1995){Nerney}, {Suess}, and {Schmahl}}]{nerney}
{Nerney}, S., {Suess}, S.~T., and {Schmahl}, E.~J.: {Flow downstream of the
  heliospheric terminal shock: Magnetic field line topology and solar cycle
  imprint}, J. of Geophys. Res., 100, 3463--3471, \doi{10.1029/94JA02690},
  1995.

\bibitem[{{Parker}(1957)}]{parker57}
{Parker}, E.~N.: {Sweet's Mechanism for Merging Magnetic Fields in Conducting
  Fluids}, J. of Geophys. Res., 62, 509--520, \doi{10.1029/JZ062i004p00509},
  1957.

\bibitem[{{Parker}(1962)}]{parker62}
{Parker}, E.~N.: Kinetic properties of interplanetary matter, Planetary and
  Space Science, 9, 461 -- 475, \doi{10.1016/0032-0633(62)90050-8},
  \urlprefix\url{http://www.sciencedirect.com/science/article/pii/003206336290%
0508}, 1962.

\bibitem[{{Parker}(1965)}]{par65}
{Parker}, E.~N.: {The passage of energetic charged particles through
  interplanetary space}, Plan. Sp. Science, 13, 9--+,
  \doi{10.1016/0032-0633(65)90131-5}, 1965.

\bibitem[{{Parker}(1970)}]{par70}
{Parker}, E.~N.: {The Generation of Magnetic Fields in Astrophysical Bodies. I.
  The Dynamo Equations}, Astrophys. J., 162, 665--+, \doi{10.1086/150697},
  1970.

\bibitem[{{Parker}(1979)}]{par79}
{Parker}, E.~N.: {Cosmical magnetic fields: Their origin and their activity},
  1979.

\bibitem[{{Petschek}(1964)}]{petschek64}
{Petschek}, H.~E.: {Magnetic Field Annihilation}, NASA Special Publication, 50,
  425, 1964.

\bibitem[{{Pogorelov} et~al.(2009{\natexlab{a}}){Pogorelov}, {Borovikov},
  {Zank}, and {Ogino}}]{pogorelov}
{Pogorelov}, N.~V., {Borovikov}, S.~N., {Zank}, G.~P., and {Ogino}, T.:
  {Three-Dimensional Features of the Outer Heliosphere Due to Coupling Between
  the Interstellar and Interplanetary Magnetic Fields. III. The Effects of
  Solar Rotation and Activity Cycle}, Astrophys. J., 696, 1478--1490,
  \doi{10.1088/0004-637X/696/2/1478}, 2009{\natexlab{a}}.

\bibitem[{{Pogorelov} et~al.(2009{\natexlab{b}}){Pogorelov}, {Heerikhuisen},
  {Zank}, {Mitchell}, and {Cairns}}]{pogorelovb}
{Pogorelov}, N.~V., {Heerikhuisen}, J., {Zank}, G.~P., {Mitchell}, J.~J., and
  {Cairns}, I.~H.: {Heliospheric asymmetries due to the action of the
  interstellar magnetic field}, Advances in Space Research, 44, 1337--1344,
  \doi{10.1016/j.asr.2009.07.019}, 2009{\natexlab{b}}.

\bibitem[{{Salvati}(2010)}]{salvati2}
{Salvati}, M.: {The local Galactic magnetic field in the direction of Geminga},
  A\&A, 513, A28+, \doi{10.1051/0004-6361/200913406}, 2010.

\bibitem[{{Salvati} and {Sacco}(2008)}]{salvati}
{Salvati}, M. and {Sacco}, B.: {The Milagro anticenter hot spots: cosmic rays
  from the Geminga supernova?}, A\&A, 485, 527--529,
  \doi{10.1051/0004-6361:200809586}, 2008.

\bibitem[{{Scherer} and {Fahr}(2003)}]{scherer03}
{Scherer}, K. and {Fahr}, H.~J.: {Breathing of heliospheric structures
  triggered by the solar-cycle activity}, Annales Geophysicae, 21, 1303--1313,
  \doi{10.5194/angeo-21-1303-2003}, 2003.

\bibitem[{{Shaikh} and {Zank}(2010)}]{zank}
{Shaikh}, D. and {Zank}, G.~P.: {Modulation of waves due to charge-exchange
  collisions in magnetized partially ionized space plasma}, Physics Letters A,
  374, 4538--4542, \doi{10.1016/j.physleta.2010.09.020}, 2010.

\bibitem[{{Shay} and {Drake}(1998)}]{shay98b}
{Shay}, M.~A. and {Drake}, J.~F.: {The role of electron dissipation on the rate
  of collisionless magnetic reconnection}, Geophys. Res. Lett., 25, 3759--3762,
  \doi{10.1029/1998GL900036}, 1998.

\bibitem[{{Shay} et~al.(1998){Shay}, {Drake}, {Denton}, and {Biskamp}}]{shay98}
{Shay}, M.~A., {Drake}, J.~F., {Denton}, R.~E., and {Biskamp}, D.: {Structure
  of the dissipation region during collisionless magnetic reconnection}, J.
  Geophys. Res., 103, 9165--9176, \doi{10.1029/97JA03528}, 1998.

\bibitem[{{Shay} et~al.(2004){Shay}, {Drake}, {Swisdak}, and {Rogers}}]{shay04}
{Shay}, M.~A., {Drake}, J.~F., {Swisdak}, M., and {Rogers}, B.~N.: {The scaling
  of embedded collisionless reconnection}, Physics of Plasmas, 11, 2199--2213,
  \doi{10.1063/1.1705650}, 2004.

\bibitem[{{Shuwang}(2011)}]{shuwang11}
{Shuwang}, C.: {Study on large-scale CR anisotropy with ARGO-YBJ experiment},
  in: International Cosmic Ray Conference, vol.~1 of {\em International Cosmic
  Ray Conference, Beijing China\/}, p. 0041, 2011.

\bibitem[{{Sturrock}(1966)}]{sturrock66}
{Sturrock}, P.~A.: {Model of the High-Energy Phase of Solar Flares}, Nature,
  211, 695--697, \doi{10.1038/211695a0}, 1966.

\bibitem[{{Sweet}(1958)}]{sweet58}
{Sweet}, P.~A.: {The Neutral Point Theory of Solar Flares}, in: Electromagnetic
  Phenomena in Cosmical Physics, edited by {B.~Lehnert}, vol.~6 of {\em IAU
  Symposium\/}, pp. 123--+, 1958.

\bibitem[{{Vernetto} et~al.(2009){Vernetto}, {Guglielmotto}, {Zhang}, and {for
  the ARGO-YBJ Collaboration}}]{vernetto}
{Vernetto}, S., {Guglielmotto}, Z., {Zhang}, J.~L., and {for the ARGO-YBJ
  Collaboration}: {Sky monitoring with ARGO-YBJ}, arXiv:0907.4615
  [astro-ph.HE], 2009.

\bibitem[{{Wefel} et~al.(2008){Wefel}, {Adams}, {Ahn}, and {et al.}}]{atic2}
{Wefel}, J.~P., {Adams}, Jr., J.~H., {Ahn}, H.~S., and {et al.}: {Revised
  Energy Spectra for Primary Elements (H - Si) above 50 GeV from the ATIC-2
  Science Flight}, in: International Cosmic Ray Conference, vol.~2 of {\em
  International Cosmic Ray Conference\/}, pp. 31--34, 2008.

\bibitem[{{Wygant} et~al.(2005){Wygant}, {Cattell}, {Lysak}, {Song}, {Dombeck},
  {McFadden}, {Mozer}, {Carlson}, {Parks}, {Lucek}, {Balogh}, {Andre}, {Reme},
  {Hesse}, and {Mouikis}}]{wygant05}
{Wygant}, J.~R., {Cattell}, C.~A., {Lysak}, R., {Song}, Y., {Dombeck}, J.,
  {McFadden}, J., {Mozer}, F.~S., {Carlson}, C.~W., {Parks}, G., {Lucek},
  E.~A., {Balogh}, A., {Andre}, M., {Reme}, H., {Hesse}, M., and {Mouikis}, C.:
  {Cluster observations of an intense normal component of the electric field at
  a thin reconnecting current sheet in the tail and its role in the shock-like
  acceleration of the ion fluid into the separatrix region}, Journal of
  Geophysical Research (Space Physics), 110, A09206,
  \doi{10.1029/2004JA010708}, 2005.

\bibitem[{{Yan} and {Lazarian}(2004)}]{yan04}
{Yan}, H. and {Lazarian}, A.: {Cosmic-Ray Scattering and Streaming in
  Compressible Magnetohydrodynamic Turbulence}, Astrophys. J., 614, 757--769,
  \doi{10.1086/423733}, 2004.

\bibitem[{{Yan} and {Lazarian}(2008)}]{yan08}
{Yan}, H. and {Lazarian}, A.: {Cosmic-Ray Propagation: Nonlinear Diffusion
  Parallel and Perpendicular to Mean Magnetic Field}, Astrophys. J., 673,
  942--953, \doi{10.1086/524771}, 2008.

\bibitem[{{Zhang}(2009)}]{argo}
{Zhang}, J.~L.: {Observation of TeV cosmic ray anisotropy by the ARGO-YBJ}, in:
  International Cosmic Ray Conference, vol.~1 of {\em International Cosmic Ray
  Conference, \L\'od\'z, Poland\/}, 2009.

\end{thebibliography}

\clearpage

\end{document}